\documentstyle[multicol,eqsecnum,aps,graphicx]{revtex}

\begin{document}
\title{Evolution of the Yukawa coupling constants and seesaw 
operators \\
 in the universal seesaw model}

\author{\bf Yoshio Koide and Hideo Fusaoka$^{(a)}$}
\address{
Department of Physics, University of Shizuoka 
52-1 Yada, Shizuoka 422-8526, Japan \\
(a) Department of Physics, Aichi Medical University  
Nagakute, Aichi 480-1195, Japan}
\date{\today}

\maketitle
\begin{abstract}
The general features of the evolution of the Yukawa coupling 
constants and seesaw operators in the universal seesaw model 
with det$M_F=0$ are investigated. In the model, not only the 
magnitude of the Yukawa coupling constant $(Y_L^u)_{33}$ in
the up-quark sector but also that of $(Y_L^d)_{33}$ in the
down-quark sector is of the order of one, i.e., 
$(Y_L^u)_{33} \sim (Y_L^d)_{33} \sim 1$.
The requirement that the model should be calculable perturbatively,
i.e., $|Y^f_{ij}|^2/4\pi \leq 1$, puts some constraints on 
the values of the intermediate mass scales and $\tan\beta$ 
(in the SUSY model).
\end{abstract}



\begin{multicols}{2}
\narrowtext

\section{Introduction}

Recently, there has been considerable interest in the
evolution (energy-scale dependency) of the Yukawa coupling 
constants of quarks and leptons.
If we intend to build a model which gives a unified 
description of quark and lepton mass matrices, we cannot
avoid investigating evolutions of the Yukawa coupling 
constants.
The recent study on the quark masses and mixings has been
given, for example, in Ref.~\cite{q-evol}.
Especially, recently, the evolution of the neutrino
seesaw mass matrix has been received considerable attention
(for example, see Ref.~\cite{nu-evol}) in connection with
the energy-scale dependence of the large mixing angle.

As one of such unified models, there is a non-standard model, 
the so-called ``universal seesaw model" (USM) \cite{USM}.
The model describes not only the neutrino
mass matrix $M_\nu$ but also the quark mass matrices
$M_u$ and $M_d$ and charged lepton mass matrix $M_e$
by the seesaw-type matrices universally:
The model has hypothetical fermions
$F_i$ ($F=U,D,N,E$; $i=1,2,3$) in addition to the conventional 
quarks and leptons $f_i$ ($f=u, d, \nu, e$; $i=1,2,3$), 
and these fermions are assigned to $f_L = (2,1)$, 
$f_R = (1,2)$, $F_L = (1,1)$ and $F_R = (1,1)$ of 
SU(2)$_L \times$ SU(2)$_R$.
The 6 $\times$ 6 mass matrix which is sandwiched
between the fields ($\overline{f}_L, \overline{F}_L$)
and ($f_R, F_R$) is given by
\begin{equation}
M^{6 \times 6} =
\left( \begin{array}{cc}
0 & m_L\\
m_R^\dagger & M_F
\end{array} \right) \ ,
\end{equation}
where $m_L$ and $m_R$ are universal for all fermion
sectors ($f=u, d, \nu, e$) and only $M_F$ have
structures dependent on the flavors $F$.
For $\Lambda_L < \Lambda_R \ll \Lambda_S$, where 
$\Lambda_L = O(m_L)$, $\Lambda_R = O(m_R)$ and
$\Lambda_S = O(M_F)$, the $3\times 3$ mass matrices
$M_f$ for the fermions $f$ are given by the
well-known seesaw expression
\begin{equation}
M_f \simeq - m_L M^{-1}_F m_R^\dagger \ .
\end{equation}
Thus, the model answers the question why the masses of 
quarks (except for top quark) and charged leptons are
so small compared with the electroweak scale
$\Lambda_L$ ($\sim$ 10$^2$ GeV).

Recently, in order to understand the observed
fact $m_t \sim \Lambda_L$ ($m_t$ is the top quark mass), 
the authors have proposed
a universal seesaw mass matrix model with an ansatz
\cite{KFzp,KFptp,Morozumi} $ {\rm det} M_F = 0$
for the up-quark sector ($F=U$).
In the model,  one of the fermion
masses $m(U_i)$ is zero [say, $m(U_3)=0$],
so that the seesaw mechanism does not work for
the third family, i.e., the fermions ($u_{3L}, U_{3R}$)
and ($U_{3L}, u_{3R}$) acquire masses of $O(m_L)$ and
$O(m_R)$, respectively. 
We identify $(u_{3L},U_{3_R})$ as the top quark 
$(t_L, t_R)$.
Thus,  we can understand the question
why only the top quark has a mass of the order of
$\Lambda_L$.

Our interest is as follows: In the conventional model, 
the Yukawa coupling constants $y_f$ of the 
fermions $f$ are given by $y_f=m_f/\langle\phi_L^0\rangle$.  
Only the Yukawa coupling constants $y_t$ of the top quark 
$t$ takes a large value $y_t=m_t/\langle\phi_L^0\rangle\sim1$.  
The other Yukawa coupling constants 
$y_f$ are sufficiently smaller than one.  
On the contrast to the conventional model, 
in this USM, the matrices $m_L^f=Y_L^f\langle\phi_L^0\rangle$ are 
universal for all fermion sectors $f=u,d,e,\nu$, i.e., 
$Y_L^u=Y_L^d=Y_L^e=Y_L^{\nu}$.  Therefore, when $(Y_L^u)_{33}$ is of the 
order of one, the other $(Y_L^f)_{33}$ will also be of the order of one.  
We are afraid that in such a model the Yukawa coupling constants
have Landau poles at energy scales lower than a unification energy
scale $\mu=\Lambda_X$ (so that the model causes ``burst" of
Yukawa coupling constants before going to the unification
energy scale).
One of our interests is to investigate whether such a model
can provide or not a set of reasonable parameter values under the
conditions that the Yukawa coupling constants  (and also 
the seesaw operators $m_L M_F^{-1} m_R^\dagger$) do not have
the Landau poles below $\mu=\Lambda_X$.

We also take an interest in the ``democratic" USM
\cite{KFzp,KFptp}, which is an extended version of
USM and has successfully given
the quark masses and the Cabibbo-Kobayashi-Maskawa
(CKM) \cite{CKM} matrix parameters in terms of the
charged lepton masses.
However, the study is only phenomenology at the
energy scale $\mu=m_Z$ ($m_Z$ is the neutral weak
boson mass).
Since the model is one of the promising models of
the unified description of the quark and lepton mass 
matrices, it is important to investigate the evolutions
of the mass matrices in the USM.

The democratic USM is as follows:

\noindent
(i) The mass matrices $m_L$ and $m_R$ have the same structure
except for their phase factors
\begin{equation}
m_R^f = \kappa m_L^f \equiv \kappa m_0 Z^f \ ,
\end{equation}
where $\kappa$ is a constant and $Z^f$ are given by
\begin{equation}
Z^f ={\rm diag}\left(z_1 \exp(i\delta_1^f),z_2 \exp(i\delta_2^f),
z_3 \exp(i\delta_3^f)\right) \ ,
\end{equation}
with $z_1^2 + z_2^2 + z_3^2 = 1$. 
(The fermion masses $m_i^f$ are independent of the parameters 
$\delta_i^f$.  Only the values of the CKM matrix parameters 
$|V_{ij}|$ depend on the parameters $\delta_i^f$).

\noindent
(ii) In the basis on which the matrices $m_L^f$ and $m_R^f$ are
diagonal, the mass matrices $M_F$ are given by the form
\begin{equation}
M_F = m_0 \lambda ({\bf 1} + 3b_f X),
\end{equation}
\begin{equation}
{\bf 1} = 
\left( \begin{array}{ccc}
1 & 0 & 0 \\
0 & 1 & 0 \\
0 & 0 & 1 \\
\end{array} \right)
,\ \ \ 
X = \frac{1}{3}
\left( \begin{array}{ccc}
1 & 1 & 1 \\
1 & 1 & 1 \\
1 & 1 & 1 \\
\end{array} \right) \ .
\end{equation}

\noindent
(iii) The parameter $b_f$ for the charged lepton sector is
given by $b_e$ = 0, so that in the limit of $\kappa/\lambda
\ll 1$, the parameters $z_i$ are given by 
\begin{equation}
\frac{z_1}{\sqrt{m_e}} = \frac{z_2}{\sqrt{m_\mu}} =
\frac{z_3}{\sqrt{m_\tau}} = \frac{1}{\sqrt{m_e + m_\mu + m_\tau}} \ .
\end{equation}
Then, the up- and down-quark masses are successfully
given by the choice of $b_u = -1/3$ and $b_d = -e^{i \beta_d}$
($\beta_d = 18^{\circ}$), respectively. The CKM matrix is also
successfully obtained by taking 
\begin{equation}
\delta_1^u -\delta_1^d=\delta_2^u-\delta_2^d=0 \ ,  \ \ 
\delta_3^u -\delta_3^d \simeq \pi \ ,
\end{equation}

However, when we take the evolution of the Yukawa coupling constants
into the consideration, we should consider that the assumptions (i) 
and (ii) are required not at the electroweak energy scale $\mu=\Lambda_L$,
but at a unification energy scale $\mu=\Lambda_X$, i.e.,
the assumptions (i) and (ii) should be replaced with

\begin{equation}
Y_L^f(\Lambda_X) = Y_R^f(\Lambda_X) = \xi_{LR}^f Z^f \ ,
\end{equation}
\begin{equation}
\xi_{LR}^u=\xi_{LR}^d,
\end{equation}
and
\begin{equation}
Y_S^f(\Lambda_X) = \xi_S^f \left( {\bf 1} + 3b_f X\right)  \ , 
\end{equation}
respectively, 
where mass matrices $m_L$, $m_R$ and $M_F$ are expressed by
\begin{equation}
m_L^f= Y_L^f \langle\phi_L^0\rangle \ , \ \ 
m_R^f= Y_R^f \langle\phi_R^0\rangle \ , \ \ 
M_F=  Y_S^f \langle\Phi^0\rangle \ , 
\end{equation}
respectively, and
\begin{equation}
\langle\phi_L^0\rangle = \langle\phi_R^0\rangle/\kappa
= \langle\Phi^0\rangle/\lambda
\end{equation}
and $\phi_L$, $\phi_R$ and $\Phi$ are Higgs scalars
whose vacuum expectation values (VEV) break  SU(2)$_L$, SU(2)$_R$, 
and an additional U(1) symmetry U(1)$_X$, respectively.
(For simplicity, we have assumed that the values of 
$\langle\phi_L^0\rangle$,  $\langle\phi_R^0\rangle$ and
$\langle\Phi^0\rangle$ are real.)

Another interest in the present paper is to check whether or not the 
phenomenological study in the previous paper \cite{KFzp} 
is still approximately valid  under the evolution of the 
Yukawa coupling constants.
For example, the model with $b_e=0$ and $b_u=-1/3$ has led to the 
relation \cite{Koide-mpl,KFzp}
\begin{equation}
{\frac{m_u}{m_c}}\simeq{\frac{3}{4}}{\frac{m_e}{m_{\mu}}},
\end{equation}
almost independently of the value of the seesaw suppression factor 
$\kappa/\lambda$.  
One of the reasons to taking the value of $b_f$ in the up-quark
sector as $b_u=-1/3$ exists in the successful relation (1.14).  
Therefore, we have interest whether the relation (1.14) still holds 
even when we take the evolution into consideration.  

Besides, even apart from such phenomenological interests, it is very
important to investigate the general features of the evolution 
of the Yukawa coupling constants in the universal seesaw model
with det$M_F=0$,
because in the present model one of the fermions $F_i$ does not
decouple from the theory at $\mu < \Lambda_S$, so that 
the evolution shows  peculiar behavior in contrast with the
conventional seesaw model.

A similar study has been done in Ref.~\cite{evol-USM} by
one of the authors (Y.K.).
However, in Ref.~\cite{evol-USM}, instead of the seesaw
operators $K^f$ which will be defined later in Eqs.~(3.8)
corresponding to $m_L M_F^{-1} m_R^\dagger$,
the evolution of the seesaw forms of the Yukawa coupling
constants $Y_L^f (Y_S^f)^{-1} (Y_R^f)^\dagger$ were
investigated by calculating the Yukawa coupling constants
$Y_L^f$, $Y_R^f$ and $Y_S^f$ individually under the 
assumption that the heavy particles with the 
masses of the order of $\Lambda_S \equiv \langle\Phi^0\rangle$
do not contribute to the evolution of $Y_A^f$ 
($A=L,R,S$)  below $\mu=\Lambda_S$.
In the present paper, we will calculate the evolution of the
Yukawa coupling constants $Y_A^f$ above $\mu=\Lambda_S$ and
that of the seesaw operators $K^f$ below $\mu=\Lambda_S$,
except for $(Y_L^f)_{i3}$ as discussed in Sec.~\ref{sec:3}.

In Sec.~\ref{sec:2}, we will discuss an additional symmetry
which is introduced for the purpose of preventing that
the fermions $F$ acquire the masses $M_F$ at the energy scale 
$\mu=\Lambda_S$.
In Sec.~\ref{sec:3}, we will give the general formulation of
the evolution of the seesaw mass matrices with ${\rm det}M_U=0$.
In Sec.~\ref{sec:4}, we give the explicit coefficients of the
renormalization group equations.
In Sec.~\ref{sec:5}, we discuss the evolution of an extended 
version of the USM, the ``democratic seesaw model" 
\cite{KFzp,KFptp}.
The numerical results for a non-SUSY model and for a minimal
SUSY model are given in Secs.~\ref{sec:6} and \ref{sec:7},
respectively.
It will be emphasized that the energy scale dependencies in the
SUSY model are quite different from those in the non-SUSY model.
The evolution of the neutrino mass matrix is given in 
Sec.~\ref{sec:8}.
It will be showed that, differently from the conventional
seesaw model, the present neutrino mass matrix is form-invariant
below $\mu=\Lambda_S$.
Finally, Sec.~\ref{sec:9} will be devoted to the conclusions and 
remarks.



\section{U(1)$_X$ symmetry}
\label{sec:2}


In the present model, the gauge symmetries are 
	broken as follows:
\begin{equation}
\begin{array}{c}
{\rm SU(2)}_L \times {\rm SU(2)}_R \times {\rm U(1)}_{LR} \times 
{\rm SU(3)}_c \times {\rm U(1)}_X \\
\downarrow \ \ \mu=\Lambda_S \\
{\rm SU(2)}_L \times {\rm SU(2)}_R \times {\rm U(1)}_{LR} \times 
{\rm SU(3)}_c \\
\downarrow \ \ \mu=\Lambda_R \\
{\rm SU(2)}_L \times  {\rm U(1)}_{Y} \times {\rm SU(3)}_c \\
\downarrow \ \ \mu=\Lambda_L \\
  {\rm U(1)}_{em} \times {\rm SU(3)}_c \ . \\
\end{array}
\end{equation}
Here, the symmetry U(1)$_X$, which is spontaneously broken 
at the energy scale $\mu=\Lambda_S$, has been introduced for the
purpose of preventing that the fermions $F$ acquire the masses
$M_F$ at $\mu > \Lambda_S$.
Hereafter, we call the ranges $\Lambda_L < \mu
\leq \Lambda_R$, $\Lambda_R < \mu \leq \Lambda_S$, and
$\Lambda_S < \mu \leq \Lambda_X$  as the ranges I, II, and
III, respectively.
In the present paper, the energy scale $\Lambda_X$ does not
always mean a gauge unification energy scale.
We assume that at the energy scale $\Lambda_X$ the mass
matrices (Yukawa coupling constants) take simple forms
discussed in the previous section.

The Yukawa coupling constants $Y_L^f$, $Y_R^f$ and $Y_S^f$ are
defined as follows:
\end{multicols}
\hspace{-0.5cm}
\rule{8.7cm}{0.1mm}\rule{0.1mm}{2mm}
\widetext
\begin{eqnarray}
H_{int} &=&
Y^u_{Lij}\overline{q}_{Li}\widetilde{\phi}_L U_{Rj}
+ Y^d_{Lij}\overline{q}_{Li}{\phi}_L D_{Rj}
+ Y^\nu_{Lij}\overline{\ell}_{Li}\widetilde{\phi}_L N_{Rj}
+ Y^e_{Lij}\overline{\ell}_{Li}{\phi}_L E_{Rj} \nonumber\\
&+& Y^u_{Rij}\overline{q}_{Ri}\widetilde{\phi}_R U_{Lj}
+ Y^d_{Rij}\overline{q}_{Ri}{\phi}_R D_{Lj}
+ Y^\nu_{Rij}\overline{\ell}_{Ri}\widetilde{\phi}_R N_{Lj}
+ Y^e_{Rij}\overline{\ell}_{Ri}{\phi}_R E_{Lj} \\
&+& Y^u_{Sij}\overline{U}_{Li}\Phi U_{Rj}
+ Y^d_{Sij}\overline{D}_{Li}{\Phi}^\dagger D_{Rj}
+ Y^\nu_{Sij}\overline{N}_{Li}{\Phi} N_{Rj}
+ Y^e_{Sij}\overline{E}_{Li}{\Phi}^\dagger E_{Rj} + h.c. \ ,\nonumber
\end{eqnarray}
\hspace{9.1cm}
\rule{-2mm}{0.1mm}\rule{8.7cm}{0.1mm}
\begin{multicols}{2}
\narrowtext
\noindent
where
\begin{eqnarray}
&&q_{L/R} =\left(
\begin{array}{c}
u \\
d
\end{array} \right)_{L/R} \ , \ \ \ 
\ell_{L/R} =\left(
\begin{array}{c}
\nu \\
e^-
\end{array} \right)_{L/R} \ ,\nonumber\\
&&\phi_{L/R} =\left(
\begin{array}{c}
\phi^+ \\
\phi^0
\end{array} \right)_{L/R} \ , \ \ \ 
\widetilde{\phi}_{L/R} =\left(
\begin{array}{c}
\overline{\phi}^0 \\
-\phi^-
\end{array} \right)_{L/R} \ .
\end{eqnarray}

{}From Eq.~(2.2), the U(1)$_X$ charge assignment should
satisfy the following relations
%
\begin{eqnarray}
&&X(U_R)= X(U_L) -X(\Phi) \ , \ \ \nonumber\\
&&X(D_R)= X(D_L) +X(\Phi) \ ,
\end{eqnarray}

\begin{eqnarray}
&&X(q_L)= \frac{1}{2} \left[ X(U_R) +X(D_R) \right] \ , \ \nonumber\\ 
&&X(q_R)= \frac{1}{2} \left[ X(U_L) +X(D_L) \right]  \ , 
\end{eqnarray}

\begin{eqnarray}
&&X(\phi_L)= \frac{1}{2} \left[ X(U_R) -X(D_R) \right]  \ , \ \nonumber\\ 
&&X(\phi_R)= \frac{1}{2} \left[ X(U_L) -X(D_L) \right]  \ , 
\end{eqnarray}
for quark sectors, and  equations similar to Eqs.~(2.4) - (2.6) 
for lepton sectors $f=\nu, e$.
For simplicity, in the present paper, we choose
\begin{equation}
X(q_{L/R}) = X(\ell_{L/R}) = 0 \ , \ \ X(\Phi)=+1 \ .
\end{equation}

Then, the quantum numbers of the fermions $f$ and $F$ and Higgs scalars 
$\phi_L$, $\phi_R$ and $\Phi$ for 
${\rm SU(2)}_L \times {\rm SU(2)}_R \times 
{\rm U(1)}_{LR} \times {\rm U(1)}_{X}$  are given in Table \ref{T-qn}.

Note that the quantum number of the fermion $N_{L}$ is identical 
with that of the fermion $N_R^c$ [$\equiv (N_R)^c \equiv 
C \overline{N}_R^T$].
Therefore, the neutral fermions $N_L$ and $N_R$ can acquire the
following Majorana mass terms at $\mu=\Lambda_S$:
\begin{equation}
H_{Majorana} = \left( Y^L_{Sij} \overline{N}_{Li} N_{Lj}^c
+ Y^R_{Sij} \overline{N}_{Ri}^c N_{Rj} \right) \Phi + h.c. \ .
\end{equation}
Then, the neutrino mass matrix is given as follows
\begin{equation}
\left( \overline{\nu}_L\ \overline{\nu}_R^c\ \overline{N}_L\ 
\overline{N}_R^c \right)
{ 
\left( \begin{array}{cccc}
0 & 0 & 0 & m_L \\
0 & 0 & m_R^{\dagger T} & 0 \\
0 & m_R^\dagger & M_L & M_D \\
m_L^T & 0 & M_D^T & M_R \\
\end{array} \right)
}
\left( \begin{array}{c}
\nu_L^c \\
\nu_R \\
N_L^c \\
N_R
\end{array} \right) \ ,
\end{equation}
where $M_D= Y_S^\nu \langle \Phi \rangle$, 
$M_L= Y_S^L \langle \Phi \rangle$ and
$M_R= Y_S^R \langle \Phi \rangle$.
Since $O(M_D) \sim O(M_L) \sim O(M_R) \gg O(m_R) \gg O(m_L)$,
we obtain the mass matrix $M_\nu$ for the active neutrinos 
$\nu_L$
\begin{equation}
M_\nu \simeq - m_L M_R^{-1} m_L^T \ .
\end{equation}



\section{General features of the evolutions}
\label{sec:3}

In the present section, we give  a general formulation of the evolution
of the seesaw matrix with det$M_F=0$.
The evolution of the neutrino seesaw mass matrix is well known.
However, in such a model with det$M_F=0$ as the present model (the
democratic seesaw model), a careful treatment is required.

Without losing the generality, we can express the Yukawa
coupling constants $Y_L^f$ and $Y_R^f$ ($f=u,d,\nu,e$)
as
\begin{equation}
Y_L^f(\mu)=\xi_L^f(\mu) Z_L^f(\mu) \ , \ \ 
Y_R^f(\mu)=\xi_R^f(\mu) Z_R^f(\mu) \ ,
\end{equation}
where $Z_A^f(\mu)$ ($A=L,R$) are defined by
\begin{equation}
Z_A^f(\mu) = {\rm diag}(z_{A1}^f(\mu), z_{A2}^f(\mu),
z_{A3}^f(\mu)) \ ,
\end{equation}
\begin{equation}
|z_{A1}^f(\mu)|^2+|z_{A2}^f(\mu)|^2+|z_{A3}^f(\mu)|^2=1 \ ,
\end{equation}
on the basis on which $Y_A^f(\mu)$ are diagonal.
In the present model, the word ``universal" means
the following initial conditions
\begin{equation}
\xi_L^f(\Lambda_X) = \xi_R^f(\Lambda_X) \equiv  \xi_{LR}
\ ,
\end{equation}
\begin{equation}
  |z_{Li}^f(\Lambda_X)|=|z_{Ri}^f(\Lambda_X)| 
\equiv z_i \ ,
\end{equation}
for all fermion sectors $f=u,d,\nu,e$ universally.

In the range III ($\Lambda_S < \mu \leq \Lambda_X$), 
the evolutions of the Yukawa coupling constants
$Y_L^f$, $Y_R^f$ and $Y_S^f$ are given by the one loop renomalization
group equations (RGE) as follows:
\begin{equation}
16\pi^2 \frac{d Y_A^f}{dt} = \left( T_A^f - G_A^f +H_A^f \right)
Y_A^f \ , \ \ (A=L, R, S) \ ,
\end{equation}
where $t=\log \mu$, and $T_A^f$, $G_A^f$ and
$H_A^f$ ($A=L, R, S$) denote contributions from fermion loop
corrections, vertex corrections due to the gauge bosons, and
vertex corrections due to the Higgs boson, respectively.
Note that the matrices $T_A^f$ and $G_A^f$ are proportional 
to the unit matrix.
As stated in the next section, the coefficients $H_A^f$ 
($A=L, R, S$) take diagonal forms on the basis on which
$Y_A^f$ are diagonal.
Therefore, if we take a basis on which $Y_L^f$ (and
$Y_R^f$) or $Y_S^f$ are diagonal at $\mu=\Lambda_X$,
then the Yukawa coupling constants $Y_L^f$ (and $Y_R^f$)
or $Y_S^f$ can keep the forms diagonal in the range III.
Sometimes, the basis on which $Y_L^f$ (and $Y_R^f$) are
diagonal is useful, but sometimes, another basis on which
$Y_S^f$ are diagonal is useful, as we discuss later.

In the present model, it is assumed that 
we can choose a flavor basis 
on which $Y_S^f(\Lambda_X)$ are 
simultaneously diagonal for all $f=u,d,\nu,e$.  
Then, on this basis, since the Yukawa coupling constants 
$Y_S^f(\mu)$ can keep the forms diagonal in the range III,
we can find that all $Y_S^f$ are diagonal at $\mu=\Lambda_S$.
We can denote those as
\begin{equation}
Y_S^f (\Lambda_S) =  {\rm diag}( y_{1S}^f, y_{2S}^f, 
y_{3S}^f)\ .
\end{equation}

At the energy scale $\mu=\Lambda_S$, the fermions $F_i$
(except for $U_3$) acquire the heavy masses 
$(M_F)_{ii}= y_{iS}^f \langle\Phi^0\rangle $.
In the conventional seesaw model with det$M_F\neq 0$, the 
energy scale behaviors of the fermion masses in $\mu<\Lambda_S$ 
are described by evolutions of the following operators
\begin{equation}
(K^f)_{ij} = \left[ Y_L^f (Y_S^f)^{-1} (Y_R^f)^{\dagger} 
\right]_{ij} = \sum_{k=1}^3 \frac{1}{y_{kS}^f} (Y_L^f)_{ik} 
(Y_R^f)_{jk}^{\ast} \ ,
\end{equation}
and
\begin{equation}
(K^\nu)_{ij} = \left[ Y_L^\nu (Y_S^\nu)^{-1} (Y_L^\nu)^T 
\right]_{ij} =\sum_{k=1}^3 \frac{1}{y_{kS}^\nu} (Y_L^\nu)_{ik} 
(Y_L^\nu)_{jk} \ .
\end{equation}
(Hereafter, for convenience, we will denote the Yukawa coupling
constants $Y_S^R$ in the Majorana mass matrix 
$M_R=Y_S^R \langle\Phi^0 \rangle$  as $Y_S^\nu$.)
The quark and lepton mass matrices $M_f$ are given by
\begin{equation}
M_f = K^f \langle\phi_L^0\rangle \langle\phi_R^0\rangle
/\langle\Phi^0\rangle \ , \  \
(f=u, d, e) \ ,
\end{equation}
\begin{equation}
M_\nu = K^\nu \langle\phi_L^0\rangle^2
/\langle\Phi^0\rangle  \ .
\end{equation}
As explicitly shown in Sec.~\ref{sec:4}, the evolutions of the operators
$K^f$ are described by the one-loop RGE's with the following
forms
\begin{eqnarray}
16\pi^2 \frac{d K^f}{dt} =  \left( T_K^f
-G_K^f \right) K^f + H_{KL}^f K^f 
+K^f H_{KR}^{f\dagger} \ ,\nonumber\\
(f=u,d,e),
\end{eqnarray}
\begin{equation}
16\pi^2 \frac{d K^\nu}{dt} =  \left( T_K^\nu
-G_K^\nu \right) K^\nu + H_{KL}^\nu K^\nu 
+K^\nu H_{KL}^{\nu T} \ ,
\end{equation}
where $T_K^f$, $G_K^f$ and ($H_{KL}^f$, 
$H_{KR}^f$) denote 
contributions from fermion loop corrections, vertex corrections 
due to the gauge bosons, and vertex corrections due to the Higgs 
bosons $\phi_L$ and $\phi_R$, respectively.

However, in the seesaw mass matrix with det$M_F=0$, since
one of the eigenvalues of $Y_S^f$ ($f=u$) is zero (say, 
$y_{3S}^u=0$), we must calculate the following operator
\begin{equation}
(K^u)_{ij} = \left[ Y_L^u (Y_S^u)^{-1} Y_R^{u\dagger} 
\right]_{ij} =\sum_{k=1}^2 \frac{1}{y_{kS}^u} (Y_L^u)_{ik} 
(Y_R^u)_{jk}^{\ast} \ ,
\end{equation}
where $Y_S^u= {\rm diag}(y_{1S}^u, y_{2S}^u)$.
Note that the matrices $Y_U$ and $Y_L^u$ ($Y_R^u$) in Eq.~(3.14) are 
$2\times 2$ and $3\times 2$ matrices, respectively.

Note that in (3.14) we have taken the sum over $k=1$ and $2$ only.
In the range II, the evolutions of the Yukawa coupling constants 
$Y_{Li3}^u$ and $Y_{R i3}^u$ ($i=1,2,3$) are still described by
the equation (3.6).
At the energy scale $\mu=\Lambda_R$, we obtain a new mass term
\begin{equation}
H_{mass}=\sum_i (Y_R^u)_{i3}^{\ast} \overline{U}_{L3} u_{Ri} 
\langle\phi_R^0\rangle \ .
\end{equation}
By defining a mixing state
\begin{equation}
u'_{R3}=\frac{(Y_R^u)_{13}^{\ast} u_{R1} +(Y_R^u)_{23}^{\ast} u_{R2} 
+(Y_R^u)_{33}^{\ast} u_{R3}}{
\sqrt{ |(Y_R^u)_{31}|^2 +|(Y_R^u)_{32}|^2 +|(Y_R^u)_{33}|^2 }} \ ,
\end{equation}
we obtain a mass $m_{t'}$ of the fourth up-quark 
$t'=(t'_L, t'_R)=(U_{L3}, u'_{R3})$,
\begin{equation}
m_{t'}= \langle\phi_R^0\rangle
\sqrt{ |(Y_R^u)_{13}|^2 +|(Y_R^u)_{23}|^2 +|(Y_R^u)_{33}|^2 } \ .
\end{equation}
Similarly, in the approximation in which the terms suppressed 
by $y_{1S}^u$ and $y_{2S}^u$ are neglected, 
the  mass $m_t$ of the third up-quark
(i.e., top quark) $t=(t_L, t_R)=(u'_{L3}, U_{R3})$ is  given by
\begin{equation}
m_t \simeq  \langle\phi_L^0\rangle
\sqrt{ |(Y_L^u)_{13}|^2 +|(Y_L^u)_{23}|^2 +|(Y_L^u)_{33}|^2 } 
 \ ,
\end{equation}
where
\begin{equation}
u'_{L3}\simeq \frac{(Y_L^u)_{13}^{\ast} u_{L1} +(Y_L^u)_{23}^{\ast} u_{L2} 
+(Y_L^u)_{33}^{\ast} u_{L3}}{
\sqrt{ |(Y_L^u)_{13}|^2 +|(Y_L^u)_{23}|^2 +|(Y_L^u)_{33}|^2 }} \ .
\end{equation}
More precisely speaking, the masses $(m_u, m_c, m_t,m_{t'})$
are obtained by diagonalizing the following mass matrix
\end{multicols}

\hspace{-0.5cm}
\rule{8.7cm}{0.1mm}\rule{0.1mm}{2mm}
\widetext
\begin{equation}
M^u = \langle\phi_L^0\rangle \left(
\begin{array}{cccc}
-(\kappa/\lambda)K_{11}^u  & 
-(\kappa/\lambda)K_{12}^u  & 
-(\kappa/\lambda)K_{13}^u  & Y_{L13}^u \\
-(\kappa/\lambda)K_{21}^u  & 
-(\kappa/\lambda)K_{22}^u  & 
-(\kappa/\lambda)K_{23}^u  & Y_{L23}^u \\
-(\kappa/\lambda)K_{31}^u  & 
-(\kappa/\lambda)K_{32}^u & 
-(\kappa/\lambda)K_{33}^u  & Y_{L33}^u \\
\kappa Y_{R13}^{u\ast} & \kappa Y_{R23}^{u\ast} & 
\kappa Y_{R33}^{u\ast} & 0 \\
\end{array} \right) \ ,
\end{equation}
\hspace{9.1cm}
\rule{-2mm}{0.1mm}\rule{8.7cm}{0.1mm}
\begin{multicols}{2}
\narrowtext
\noindent
which is sandwiched by the fields $(\overline{u}_{L1},
\overline{u}_{L2}, \overline{u}_{L3}, \overline{U}_{L3})$ and
$({u}_{R1},{u}_{R2}, {u}_{R3},{U}_{R3})$, where $\kappa=
\langle\phi_R^0\rangle/\langle\phi_L^0\rangle$ and
$\lambda=\langle\Phi^0\rangle/\langle\phi_L^0\rangle$ 
as defined in Eq.~(1.13).
Of the Yukawa coupling constants $(Y_L^u)_{ij}$ and $(Y_R^u)_{ij}$,
the twelve components $(Y_L^u)_{ik}$ and $(Y_R^u)_{ik}$ ($i=1,2,3;
j=1,2$) are absorbed into the operator $K^u$ defined by
(3.14), while  the rest $(Y_L^u)_{i3}$ and $(Y_R^u)_{i3}$ are still
described by the equation (3.6).

Finally, we denote the effective Hamiltonian in each range:
The effective Hamiltonian $H^{III}_{int}$ in the range III
($\Lambda_X \geq \mu > \Lambda_S$) is still given by the form (2.2),
and $H^{II}_{int}$ in the range II ($\Lambda_S \geq \mu > \Lambda_R$) 
and $H^{I}_{int}$ in the range I ($\Lambda_R \geq \mu > \Lambda_L$)
are given by
\begin{eqnarray}
H_{int}^{II} &=&
\sum_{i=1}^3 Y^u_{Li3}(\overline{q}_{Li}\widetilde{\phi}_L U_{R3})
\nonumber \\
&+& \sum_{i=3}^3 Y^u_{Ri3}(\overline{q}_{Ri}\widetilde{\phi}_R 
U_{L3}) \nonumber \\
&+& \sum_{i,j\neq 3}\frac{1}{\langle \Phi^0\rangle }
K^u_{ij} (\overline{q}_{Li} \widetilde{\phi}_L )
(\widetilde{\phi}_R^\dagger q_{Rj}) \nonumber \\
&+& \sum_{i,j}\frac{1}{\langle \Phi^0\rangle }
K^d_{ij} (\overline{q}_{Li} {\phi}_L )
({\phi}_R^\dagger q_{Rj}) \nonumber \\
&+& \sum_{i,j}\frac{1}{\langle \Phi^0\rangle }
K^e_{ij} (\overline{\ell}_{Li} {\phi}_L )
({\phi}_R^\dagger \ell_{Rj}) +h.c. \nonumber \\
&+& \sum_{i,j}\frac{1}{\langle \Phi^0\rangle }
K^\nu_{L ij} (\overline{\ell}_{Li} \widetilde{\phi}_L )
(\widetilde{\phi}_L^T \ell_{Lj}^c) \nonumber \\
&+& \sum_{i,j}\frac{1}{\langle \Phi^0\rangle }
K^\nu_{R ij} (\overline{\ell}_{Ri} \widetilde{\phi}_R )
(\widetilde{\phi}_R^T \ell_{Rj}^c) \ , 
\end{eqnarray}
and
\begin{eqnarray}
H_{int}^{I} &=&
\sum_{i=1}^3 Y^u_{Li3}(\overline{q}_{Li}\widetilde{\phi}_L U_{R3})
\nonumber \\
&+& \sum_{i,j\neq 3}\frac{\langle \phi_R^0\rangle}{\langle \Phi^0\rangle }
K^u_{ij} (\overline{q}_{Li} \widetilde{\phi}_L u_{Rj}) \nonumber \\
&+& \sum_{i,j}\frac{\langle \phi_R^0\rangle}{\langle \Phi^0\rangle }
K^d_{ij} (\overline{q}_{Li} {\phi}_L d_{Rj}) \nonumber \\
&+& \sum_{i,j}\frac{\langle \phi_R^0\rangle}{\langle \Phi^0\rangle }
K^e_{ij} (\overline{\ell}_{Li} {\phi}_L e_{Rj}) +h.c. \nonumber \\
&+& \sum_{i,j}\frac{1}{\langle \Phi^0\rangle }
K^\nu_{L ij} (\overline{\ell}_{Li} \widetilde{\phi}_L )
(\widetilde{\phi}_L^T \ell_{Lj}^c) \ ,
\end{eqnarray}
\noindent
respectively.



\section{Coefficients of the RGE}
\label{sec:4}


In the present section, we give the coefficients of the renomalization group
equations (RGE) (3.6), (3.12) and (3.13).

\subsection{Evolution in the range III}

In the non-SUSY model, the terms $T_A^f$, $G_A^f$ and $H_A^f$ 
($A=L,R,S$) are given as follows:

\begin{eqnarray}
 T_A^u&=&T_A^d=T_A^\nu=T_A^e  \nonumber\\
 &=&3 {\rm Tr}\left( Y_A^u Y_A^{u \dagger} + Y_A^d Y_A^{d \dagger}\right)
+ {\rm Tr}\left( Y_A^\nu Y_A^{\nu \dagger} + Y_A^e Y_A^{e \dagger}
\right) \ , \nonumber \\
\end{eqnarray}
\begin{eqnarray}
&&G_A^u= \frac{17}{8}g_1^2 +\frac{9}{4}g_{2A}^2 +8 g_3^2
+\frac{3}{4} g_X^2  \ , \ \ \ \nonumber\\
&&G_A^d= \frac{5}{8}g_1^2 +\frac{9}{4}g_{2A}^2 +8 g_3^2 
+\frac{3}{4} g_X^2  \ , \nonumber \\
&&G_A^\nu= \frac{9}{8}g_1^2 +\frac{9}{4}g_{2A}^2
+\frac{3}{4} g_X^2  \ , \ \ \nonumber\\ 
&&G_A^e= \frac{45}{8}g_1^2 +\frac{9}{4}g_{2A}^2
+\frac{3}{4} g_X^2  \ ,
\end{eqnarray}
\begin{eqnarray}
&&H_A^u=-H_A^d =\frac{3}{2}\left( Y_A^u Y_A^{u \dagger} 
- Y_A^d Y_A^{d \dagger} \right) \ ,\nonumber\\
&&H_A^\nu=-H_A^e =\frac{3}{2}\left( Y_A^\nu Y_A^{\nu \dagger} - 
Y_A^e Y_A^{e \dagger} \right) \ ,
\end{eqnarray}
where $A=L,R$, and 
\begin{eqnarray}
T_S^u&=&T_S^d=T_S^\nu=T_S^e \nonumber\\
&=&3 {\rm Tr}\left( Y_S^u Y_S^{u \dagger} + Y_S^d Y_S^{d \dagger}\right)
+ {\rm Tr}\left( Y_S^\nu Y_S^{\nu \dagger} + Y_S^e Y_S^{e \dagger}
\right) \ , \nonumber\\
\end{eqnarray}
\begin{eqnarray}
&&G_S^u= 4 g_1^2  +8 g_3^2 +\frac{3}{2} g_X^2 \ , \ \ \ \nonumber\\
&&G_S^d=  g_1^2  +8 g_3^2 +\frac{3}{2}  g_X^2  \ , \nonumber \\
&&G_S^\nu= +\frac{3}{2} g_X^2  \ , \ \ \nonumber\\ 
&&G_S^e= 9 g_1^2 +\frac{3}{2} g_X^2  \ , 
\end{eqnarray}
\begin{equation}
H_S^f= Y_S^f Y_S^{f \dagger} \ , \ \ \ \ \ (f=u,d,\nu,e) \ .
\end{equation}

The coefficients $T_A^f$, $G_A^f$ and $H_A^f$ in the minimal SUSY
model are given in the Appendix.

As seen from Eq.~(4.6), since the matrix $H_A^f$ is diagonal on the
diagonal basis of $M_F(\Lambda_X)$, the Yukawa coupling constants 
$Y_S^f(\mu)$ can keep the forms diagonal.
Similarly, when we choose the diagonal basis of $M_L(\Lambda_X)$
[$M_R(\Lambda_X)$], the matrices $Y_L^f(\mu)$ [$Y_R^f(\mu)$] keep 
their forms diagonal.

For a model with $g_{2L}(\mu)=g_{2R}(\mu)$ 
and $Y_L^f (\mu)=Y_L^f (\mu)$ at $\mu=\Lambda_X$, we can assert that
\begin{equation}
Y_L^f(\mu) = Y_R^f(\mu) \ , 
\end{equation}
in the range III ($\Lambda_S < \mu \leq \Lambda_X$),
because on the diagonal basis of $Y_L$ we obtain
\begin{equation}
16 \pi^2 \frac{d}{dt} \ln \frac{(Y_L^f)_{ii}}{(Y_R^f)_{ii}}
=(T_L^f-G_L^f+H_L^f)_{ii} -(T_R^f-G_R^f+H_R^f)_{ii} \ .
\end{equation}
The case $g_{2L}=g_{2R}$ is likely in the L-R symmetric model.
For convenience, in the numerical evaluation in the present paper,
we will  take $g_{2L}(\mu)=g_{2R}(\mu)$ in the range III 
($\Lambda_S < \mu \leq \Lambda_X$).

\subsection{Evolution in the ranges I and II}


In the ranges I and II, all the fermions $F_i$ except for $U_3$ are
decoupled from the equation (3.6). 
In the present section, we will take the diagonal basis of $M_F$.
Therefore, it is convenient that we define a spurion
\begin{equation}
S=\left(
\begin{array}{ccc}
0 & 0 & 0 \\
0 & 0 & 0 \\
0 & 0 & 1
\end{array} \right) \ .
\end{equation}
Then, the surviving Yukawa coupling constants $(Y^u_A)_{i3}$ are
expressed as $(Y^u_{A} S)_{ij} =(Y^u_A)_{i3}\delta_{3j}$.
The evolution of $Y^u_AS$ is still described by the RGE (3.6)
by substituting $Y_A^u S$ for $Y_A^f$.
Here, the terms $T_A^u$, $G_A^u$ and $H_A^u$ ($A=L,R$) are 
expressed as follows [$Y_A^f$ ($f=d,e,\nu$) are already absorbed 
into the operators $K^f$]:
\begin{equation}
T_A^u= 3 {\rm Tr}\left( Y_A^u  S Y_A^{u\dagger} \right) \ ,
\end{equation}
\begin{equation}
G_A^u =  \frac{17}{8}g_1^2 +\frac{9}{4}g_{2A}^2 +8g_3^2  \ ,
\end{equation}
\begin{equation}
H_A^u = \frac{3}{2}Y_A^u S Y_A^{u\dagger} \ ,
\end{equation}
($A=L,R$) in the range II, and
\begin{equation}
T_L^u= 3 {\rm Tr}\left( Y_L^u  S Y_L^{u \dagger} \right) \ ,
\end{equation}
\begin{equation}
G_L^u =  \frac{17}{20}g_1^2 +\frac{9}{4}g_{2L}^2 +8g_3^2  \ ,
\end{equation}
\begin{equation}
H_L^u = \frac{3}{2}Y_L^u S Y_L^{u\dagger} \ , 
\end{equation}
in the range I.
Here, the coupling constant $g_1\equiv g_{1LR}$ 
in the range II is that for the U(1) 
operator $(1/2)Y_{LR}$ which is defined by the relation
\begin{equation}
Q=I_3^L + I_3^R +\frac{1}{2} Y_{LR} \ ,
\end{equation}
for the symmetry 
${\rm SU(2)}_L \times {\rm SU(2)}_R \times {\rm U(1)}_{LR}$, while
the coupling constant $g_1\equiv g_{1Y}$ 
in the range I is that for the U(1) 
operator $(1/2)Y$ which is defined by the relation
\begin{equation}
Q=I_3^L +\frac{1}{2} Y \ ,
\end{equation}
for the symmetry ${\rm SU(2)}_L \times {\rm U(1)}_{Y}$,
and they are connected by
\begin{equation}
\alpha_{em}^{-1}(\Lambda_L)= \alpha_{2L}^{-1}(\Lambda_L) 
+\frac{5}{3} \alpha_{1LR}^{-1}(\Lambda_L) \ ,
\end{equation}
\begin{equation}
\frac{5}{3} \alpha_{1Y}^{-1}(\Lambda_R) = 
\alpha_{2R}^{-1}(\Lambda_R) 
+\frac{2}{3} \alpha_{1LR}^{-1}(\Lambda_R) \ ,
\end{equation}
where $\alpha_i=g_i^2/4\pi$.

Similarly, the terms $T_K^f$, $G_K^f$ $H_{KL}^f$ and 
$H_{KR}^f$ ($f=u,d,e$) are given by
\begin{equation}
T_K^u=T_K^d=T_K^e=3{\rm Tr}\left( Y_L^u S Y_L^{u\dagger}
+Y_R^u S Y_R^{u\dagger} \right) \ ,
\end{equation}
\begin{eqnarray}
G_K^u&=&G_K^d=\frac{5}{2}g_1^2 +\frac{9}{4}g_{2L}^2 
+\frac{9}{4}g_{2R}^2 + 8 g_3^2 \ ,  \nonumber \\
G_K^e&=&\frac{9}{2}g_1^2 +\frac{9}{4}g_{2L}^2 
+\frac{9}{4}g_{2R}^2  \ ,
\end{eqnarray}
\begin{equation}
H_{KA}^u=H_{KA}^d=\frac{3}{2} Y_A^{u} S Y_A^{u\dagger}\ , 
\ \  H_{KA}^e =0\ , \ \ (A=L,R) \ ,
\end{equation}
in the range II, and
\begin{equation}
T_K^u=T_K^d=T_K^e=3{\rm Tr}\left( Y_L^u S Y_L^{u\dagger}
 \right) \ ,
\end{equation}
\begin{eqnarray}
G_K^u&=&\frac{17}{20}g_1^2 +\frac{9}{4}g_{2L}^2 
 + 8 g_3^2 \ , \nonumber \\
G_K^d&=&\frac{5}{20}g_1^2 +\frac{9}{4}g_{2L}^2 
 + 8 g_3^2 \ ,\\
G_K^e&=&\frac{45}{20}g_1^2 +\frac{9}{4}g_{2L}^2 
 \ , \nonumber
\end{eqnarray}
\begin{equation}
H_{KL}^u= H_{KL}^d= \frac{3}{2} Y_L^{u} S Y_L^{u\dagger}\ , 
\ \ H_{KR}^f=0 \ , \ 
\ f=u,d \ ,
\end{equation}
\begin{equation}
H_{KL}^e= H_{KR}^e=0 \ , 
\end{equation}
in the range I.

The terms $T_K^\nu$, $G_K^\nu$ and $H_{KL}^\nu$ have rather
simple forms in contrast with those in the conventional neutrino
seesaw model, because the partners of the fermions $f_L$ which couple
to the Higgs scalar $\phi_L$ are not $f_R$, but $F_R$ which are already
decoupled at $\mu=\Lambda_R$:
\begin{equation}
T_K^\nu=6{\rm Tr}\left( Y_L^u S Y_L^{u\dagger} \right) \ ,
\end{equation}
\begin{equation}
G_K^\nu= 3 g_{2L}^2  \ ,
\end{equation}
\begin{equation}
H_{KL}^\nu= \lambda_{HL} \ ,
\end{equation}
in the ranges I and II, 
where $\lambda_{HL}$ is the coupling constant of the Higgs scalar
$\phi_L$ defined by
\begin{equation}
H_\phi = \frac{1}{2} \lambda_{HL} (\phi_L^\dagger \phi_L)^2 \ ,
\end{equation}
and the mass of the physical Higgs scalar $H_L^0$ is given by
\begin{equation}
m_{HL}^2 = 2 \lambda_{HL} \langle \phi_L^0 \rangle^2 \ .
\end{equation}

The similar coefficients in the minimal SUSY model are given in the
Appendix.


\section{Case of the democratic seesaw model}
\label{sec:5}


In the democratic seesaw model, on the diagonal basis of 
$Y_L^f(\Lambda_X)$ and $Y_R^f(\Lambda_X)$, the Yukawa
coupling constants of heavy fermions $Y_S^f(\Lambda_X)$ are
given by the democratic form (1.11).
Since on this basis the Yukawa coupling constants $Y_S^f$ keep
the forms democratic:
\begin{equation}
Y_S^f(\mu) = \xi_S^f(\mu) \left( {\bf 1} + 
3b_f(\mu) {X}\right)  \ , 
\end{equation}
we will
call this basis the ``democratic basis of $M_F$" hereafter.
On the other hand, if we take a basis on which $Y_S^f$ 
are diagonal, i.e., the matrix forms are given by
\begin{equation}
\widetilde{Y}_S^f(\mu) = \xi_S^f(\mu) \left( {\bf 1} + 3b_f(\mu) 
\widetilde{X} \right)  \ , 
\end{equation}
\begin{equation}
\widetilde{X} = A X A^T ={\rm diag}(0, 0, 1) \ ,
\end{equation}
\begin{equation}
A= \left(
\begin{array}{ccc}
\frac{1}{\sqrt{2}} & - \frac{1}{\sqrt{2}} & 0 \\
\frac{1}{\sqrt{6}} & \frac{1}{\sqrt{6}} & 
- \frac{2}{\sqrt{6}} \\
 \frac{1}{\sqrt{3}} & \frac{1}{\sqrt{3}} & 
 \frac{1}{\sqrt{3}} \\
\end{array} \right) \ .
\end{equation}
Especially, on this basis, the Yukawa coupling constants 
$(\widetilde{Y}_S^e)_{ii}$ and $(\widetilde{Y}_S^u)_{ii}$  
of the fermions $E_i$ and $U_i$ satisfy the relations
\begin{equation}
[\widetilde{Y}_S^e(\mu)]_{11}= [\widetilde{Y}_S^e(\mu)]_{22} 
=[\widetilde{Y}_S^e(\mu)]_{33} = \xi_S^e(\mu)  \ , 
\end{equation}
\begin{equation}
[\widetilde{Y}_S^u(\mu)]_{11}= [\widetilde{Y}_S^u(\mu)]_{22}=
\xi_S^u(\mu)  \ ,
\ \ \  [\widetilde{Y}_S^u(\mu)]_{33}=0 \ ,
\end{equation}
in the range III ($\Lambda_S < \mu \leq \Lambda_X$), i.e.,
\begin{equation}
b_e(\mu)=0 \ , \ \ \ \ b_u(\mu)=-1/3 \ .
\end{equation}
On the other hand, on this basis, 
the Yukawa coupling constants 
$\widetilde{Y}_L^f(\mu)$ and $\widetilde{Y}_R^f(\mu)$ are not diagonal. 
However, we can easily obtain their diagonal forms by 
$A^T \widetilde{Y}_L^f(\mu) A$ and $A^T \widetilde{Y}_R^f(\mu) A$. 

At the energy scale $\mu=\Lambda_S$, the fermions $F_i$
(except for $U_3$) acquire the heavy masses $(M_F)_{ii}$.
Therefore, for $\mu < \Lambda_S$, the operators $K^f$ are given 
as follows:
\end{multicols}
\widetext
\begin{equation}
(K^u)_{ij} = \left[ \widetilde{Y}_L^u (\widetilde{Y}_S^u)^{-1} 
\widetilde{Y}_R^{u\dagger} \right]_{ij}
=\frac{1}{\xi^u_S(\Lambda_S)}
\sum_{k=1,2} (\widetilde{Y}_L^u)_{ik} (\widetilde{Y}_R^u)_{jk}^{\ast} \ ,
\end{equation}
\begin{equation}
(K^d)_{ij} = \left[ \widetilde{Y}_L^d (\widetilde{Y}_S^d)^{-1} 
\widetilde{Y}_R^{d\dagger} \right]_{ij}
=\frac{1}{\xi^d_S(\Lambda_S)}\left( \sum_{k=1,2} (\widetilde{Y}_L^d)_{ik} 
(\widetilde{Y}_R^d)_{jk}^{\ast} 
+\frac{1}{1+3b_d(\Lambda_S)} (\widetilde{Y}_L^d)_{i3} 
(\widetilde{Y}_R^d)_{j3}^{\ast} \right) \ ,
\end{equation}
\begin{equation}
(K^e)_{ij} = \left[ \widetilde{Y}_L^e (\widetilde{Y}_S^e)^{-1} 
\widetilde{Y}_R^{e\dagger} \right]_{ij}
=\frac{1}{\xi_S^e(\Lambda_S)} 
\sum_{k=1,2,3} (\widetilde{Y}_L^e)_{ik} (\widetilde{Y}_R^e)_{jk}^{\ast} 
\ ,
\end{equation}
\begin{equation}
(K^\nu)_{ij} = \left( \widetilde{Y}_L^\nu (\widetilde{Y}_S^\nu)^{ -1} 
\widetilde{Y}_L^{\nu T} \right)_{ij}
=\frac{1}{\xi_S^\nu(\Lambda_S)}
\left( \sum_{k=1,2} (\widetilde{Y}_L^\nu)_{ik} (\widetilde{Y}_L^\nu)_{jk} 
+\frac{1}{1+3b_\nu(\Lambda_S)} (\widetilde{Y}_L^\nu)_{i3} 
(\widetilde{Y}_L^\nu)_{j3} \right) \ .
\end{equation}
\hspace{9.1cm}
\rule{-2mm}{0.1mm}\rule{8.7cm}{0.1mm}
\begin{multicols}{2}
\narrowtext
\noindent
In Eq.~(5.11), we have assumed that the structure of the Majorana 
mass term $M_R(\Lambda_S)= Y_S^R(\Lambda_S) \langle\Phi^0\rangle$  
for the neutral fermions $N_R$ has  a structure similar to 
the Dirac mass matrices $M_F$ which is given by Eq.~(5.1).

Since the Yukawa coupling constants $Y_A^f(\mu)$ ($A=L,R$) in the
range III keep their forms diagonal on the democratic basis of 
$M_F$, it is convenient to express $Y_A^f(\mu)$ as follows,
\begin{equation}
Y_A^f(\mu) = \xi_A^f (\mu) Z_A^f (\mu) \ ,
\end{equation}
where the diagonal matrix $Z_A^f(\mu)$ is given by Eq.~(3.2).
Then, the matrix $\widetilde{Y}_A^f$ on the diagonal basis of $M_F$
is given by
\begin{equation}
\widetilde{Y}_A^f(\mu) = \xi_A^f(\mu) \widetilde{Z}^f (\mu) \ ,
\end{equation}
where
\end{multicols}
\hspace{-0.5cm}
\rule{8.7cm}{0.1mm}\rule{0.1mm}{2mm}
\widetext
\begin{equation}
 \widetilde{Z}^f = A Z^f A^T = \frac{1}{6} \left(
\begin{array}{ccc}
3(z_2+z_1) & -\sqrt{3} (z_2-z_1) & -\sqrt{6} (z_2-z_1) \\
-\sqrt{3} (z_2-z_1) & 4z_3 +z_2+z_1 & -\sqrt{2} (2z_3-z_2-z_1) \\
-\sqrt{6} (z_2-z_1) & -\sqrt{2} (2z_3-z_2-z_1) & 2(z_3+z_2+z_1)
\end{array} \right) \ ,
\end{equation}
\hspace{9.1cm}
\rule{-2mm}{0.1mm}\rule{8.7cm}{0.1mm}
\begin{multicols}{2}
\narrowtext
\noindent
(we have dropped the indices $A$ and $f$, and for simplicity, we
have taken $\delta_i^f=0$).
Although the Yukawa coupling constants $\widetilde{Y}_L^u$ and 
$\widetilde{Y}_R^u$ in the range II and $\widetilde{Y}_L^f$ 
in the range I have the physical meaning only 
for the one column matrix components $(\widetilde{Y}_A^f)_{i3}$ 
($i=1,2,3$), we still use the expressions (5.12) and (5.13), 
because the matrix
$K^e(\mu)$ ($\Lambda_L <\mu \leq \Lambda_S$) which is proportional
to $Y_L^e(\mu) Y_R^{e\dagger}(\mu)$ is still diagonal on the
democratic basis of $M_F$ as discussed in Sec.~\ref{sec:4}, so
that we regard that $Y_A^u(\mu)$ is also ``diagonal".
Then, the top quark mass $m_t(\mu)$ is approximately expressed as
\end{multicols}
\hspace{-0.5cm}
\rule{8.7cm}{0.1mm}\rule{0.1mm}{2mm}
\widetext
\begin{equation}
m_t(\mu) \simeq  \langle\phi_L^0\rangle
\sqrt{ \sum _i |(\widetilde{Y}_L^u(\mu))_{i3}|^2}
= \langle\phi_L^0\rangle \xi_L^u(\mu) \sqrt{ \frac{1}{3}
\sum_i |z_{Li}^u|^2 } = \frac{1}{\sqrt{3}} \xi_L^u(\mu) 
\langle\phi_L^0\rangle 
 \ .
\end{equation}
\hspace{9.1cm}
\rule{-2mm}{0.1mm}\rule{8.7cm}{0.1mm}
\begin{multicols}{2}
\narrowtext
\noindent
The expression (5.15) is valid in the whole ranges 
$\Lambda_L < \mu \leq \Lambda_X$.

Since
\begin{equation}
\sum_{i=1}^3 \sum_{k=1}^2 \left(\widetilde{Z}_{ik}\right)^2 
=\frac{2}{3} (z_1^2+z_2^2+z_3^2) = \frac{2}{3} \ ,
\end{equation}
we obtain
\begin{equation}
m_c(\mu)+m_u(\mu) \simeq \frac{2}{3} \frac{\xi^u_L(\mu) \xi^u_R(\mu)}
{\xi^u_S(\mu)} \frac{\langle\phi^0_L\rangle \langle\phi^0_R\rangle}
{\langle\Phi^0\rangle} \ ,
\end{equation}
from Eq.~(5.8).
Note that the expression (5.17) is valid only in the range III.
In the ranges I and II, the ratio $\xi^u_L \xi^u_R/\xi^u_S$
behaves as a operator $K^u(\mu)$ which obeys Eq.~(3.12).
{}From Eq.~(5.17), the ratio $m_c/m_t$ is given by
\begin{equation}
\frac{m_c(\mu)}{m_t(\mu)} \simeq \frac{2}{\sqrt{3}} 
\frac{\xi^u_R(\mu)}{\xi^u_S(\mu)} 
\frac{\langle\phi^0_R\rangle}{\langle\Phi^0\rangle} \ .
\end{equation}

Since $H^e_{KL}=H^e_{KR}=0$ in the ranges I and II, the form of
$K^e(\mu)$ is invariant in the ranges, i.e., 
\begin{equation}
Z^e_L(\Lambda_L) Z^{e\dagger}_R(\Lambda_L) = 
Z^e_L(\Lambda_S) Z^{e\dagger}_R(\Lambda_S) \ ,
\end{equation}
especially, since 
\begin{equation}
Z^f_L(\mu) = Z_R^f(\mu) \equiv Z^f(\mu) \ ,
\end{equation}
for a model with $g_{2R}(\Lambda_R)=g_{2L}(\Lambda_R)$,
we obtain
\begin{equation}
Z^e(\Lambda_L) = Z^e(\Lambda_S) \ .
\end{equation}
Therefore, in preliminary evaluations prior to fixing the final values
of the parameters, we will sometimes use the values of $z_i(m_Z)$ 
which are obtained from the observed charge lepton masses 
$m^e_i(m_Z)$ by using Eq.~(1.7) instead of the values of $z_i(\Lambda_X)$
which are defined in Eq.~(1.9) as the initial condition at $\mu=\Lambda_X$.



\section{Numerical results in the non-SUSY model}
\label{sec:6}


We define
\begin{equation}
\Lambda_L =\langle\phi_L^0\rangle \ , \ \ 
\Lambda_R =\langle\phi_R^0\rangle \ , \ \ 
\Lambda_S =\langle\Phi^0\rangle \ .  
\end{equation}
However, for convenience, in the numerical evaluations, 
instead of physical quantities at $\mu=\Lambda_L$,
we will use those at $\mu=m_Z$ ($m_Z$ is the neutral weak
boson mass).

First, in order to overlook the behavior of the Yukawa coupling 
constant $Y_L^f(\mu)$, we illustrate the behavior of 
$\xi_L^u(\mu)$ in the non-SUSY model in Fig.~\ref{xi-nonSUSY}.
Here, we have used the approximate relation (5.15) and the input 
values 
$m_t(m_Z)=181$ GeV and $\langle \phi_L^0 \rangle = 174$ GeV:
\begin{equation}
\xi_L^u(m_Z)={\sqrt{3}}
{\frac{m_t(m_Z)}{\langle\phi_L^0\rangle}}=1.80.
\end{equation}
In other words, the behavior of $\xi^u_L(\mu)$ corresponds to that of
$m_t(\mu)$ because of $\xi^u_L(\mu)=(m_t(\mu)/m_t(m_Z))\xi^u_L(m_Z)$.
In the ranges I and II, since the terms $T_L^u$ and $H_L^u$ are  
expressed only in terms of $\widetilde{Y}_L^u S\widetilde{Y}_L^{u\dagger}$,
the evolution of the factor $|\xi_L^u(\mu)|^2= 3 {\rm Tr}[ 
\widetilde{Y}_L^u S \widetilde{Y}_L^{u\dagger} ]$ is described by the
equation
\begin{equation}
16\pi^2 \frac{d}{dt} |\xi_L^u|^2 =2 \left[ \left( \frac{1}{3}
|\xi_L^u|^2 -G_L^u\right) |\xi_L^u|^2 + \frac{1}{2} |\xi_L^u|^4
\right] \ .
\end{equation}
However, in the range III, the terms $T_L^u$ and $H_L^u$ contain
other factors $Y_L^f Y_L^{f\dagger}$ in addition to 
$Y_L^u Y_L^{u\dagger}$, so that the evolution of $\xi_L^u$ cannot be
expressed so simply such as (6.3).
For the evaluation of $\xi_L^u$ in the range III, 
we have tentatively substituted the values $z_i(m_Z)$ given by (1.7) 
for the initial values $z_i(\Lambda_X)$.
For simplicity, as we discussed in (4.7), we have taken as
$g_{2L}(\Lambda_R) =g_{2R}(\Lambda_R)$.
In Fig.~\ref{xi-nonSUSY}, the ratio $\Lambda_S/\Lambda_R$ has been 
taken as $\Lambda_S/\Lambda_R=107$, which has determined 
from the fitting of the observed ratio $m_t/m_c$ as we discuss later.
The  behavior of $\xi^u_L(\mu)$ is insensitive to the ratio
$\Lambda_S/\Lambda_R$.  
As seen in Fig.~\ref{xi-nonSUSY}, in a case with  a lower $\Lambda_S$ 
($\Lambda_S < 10^5$ GeV), $\xi^u_L(\mu)$ has the Landau pole
below $\mu=\Lambda_X$,
so that the case is ruled out.
On the other hand, a case with a higher $\Lambda_S$ 
($\Lambda_S > 10^{19}$ GeV) causes $\alpha_1(\mu)\rightarrow
\infty$ at $\mu\rightarrow \Lambda_S$,
so that the case is also ruled out.

Taking account of the behavior of $\xi^u_L(\mu)$ shown in
Fig.~\ref{xi-nonSUSY}, as a trial, we take
\begin{equation}
\Lambda_X = 2\times 10^{16} \ {\rm GeV} \ ,
\end{equation}
which is known as the unification energy scale in the
minimal SUSY model.
(However, in the present paper, we do not consider the
gauge unification.)
As a value of $\Lambda_S$, we tentatively take
\begin{equation}
\Lambda_S = 3\times 10^{13} \ {\rm GeV} \ ,
\end{equation}
which leads to the 
mass-squared difference 
$\Delta{m}_{32}^2\equiv{m}_{{\nu}_3}^2-m_{{\nu}_2}^2\sim
(10^{-3}-10^{-2}){\rm eV}^2$ as we demonstrate later.
For the values (6.4) and (6.5), we obtain 
$\xi^u_L(\Lambda_X)=1.2$.

Next, we determine the values of $\xi^u_S(\Lambda)$ and
$\Lambda_S/\Lambda_R$.
Since we have already obtained the value $\xi^u_L(\Lambda_X)=1.2$,
it seems that we can fix the value of $\xi^u_S(\Lambda)
\Lambda_S/\Lambda_R$ from the observed value of $m_t(m_Z)/m_c(m_Z)$
because of the relation (5.18).
However, the value of $\xi^u_S(\Lambda_X)$ [also $\xi^f_S(\Lambda_X)$]
is sensitive to the value of $\xi^u_S(\Lambda_S)$ 
[$\xi^f_S(\Lambda_S)$] [in other words, a small deviation of
$\xi^f_S(\Lambda_S)$ causes a large deviation of $\xi^f_S(\Lambda_X)$].
Therefore, we cannot fix the values $\xi^u_S(\Lambda_X)$
unless we put a tentative model for $\xi^f_{LR}$ and $\xi^f_S$.
The basic assumption in the universal seesaw model is to consider
that the mass matrices $m_L$ and $m_R$ in Eq.~(1.1) are
``universal" (common) for all fermion sectors (quarks and leptons).
Therefore, we put the following initial condition
\begin{equation}
\xi^u_{LR}(\Lambda_X)=\xi^d_{LR}(\Lambda_X)=\xi^e_{LR}(\Lambda_X)=
\xi^\nu_{LR}(\Lambda_X)\equiv \xi_{LR}(\Lambda_X) \ .
\end{equation}
Then, a model with $\xi^u_S(\Lambda_X)=\xi^d_S(\Lambda_X)=
\xi^e_S(\Lambda_X)$ is obviously ruled out because we cannot
give the observed values of quark and charged lepton masses 
simultaneously.
We must consider
\begin{equation}
\xi^u_S(\Lambda_X)=\xi^d_S(\Lambda_X) \equiv \xi^q_S(\Lambda_X)
\neq \xi^e_S(\Lambda_X) \ .
\end{equation}
We tentatively put $\xi^e_S(\Lambda_X)=\xi_{LR}(\Lambda_X)$.
The numerical results are as follows:
\begin{equation}
\xi_{LR}(\Lambda_X)=\xi^e_S(\Lambda_X)=1.20 \ , \ \ \ 
\xi^q_S(\Lambda_X)=0.80 \ ,
\end{equation}
\begin{equation}
\Lambda_S/\Lambda_R = 107 \ ,
\end{equation}
\begin{equation}
z_1=0.01617\ , \ \ z_2=0.2349\ , \ \ 
z_3=0.9719\ .
\end{equation}
In the quark and charged lepton mass expressions (3.19) 
the factors $\xi^e_S$ and $\xi^q_S$  appear only in terms of the
combinations $\xi^q_S \Lambda_S$ and $\xi_S^e \Lambda_S$, respectively,
so that the absolute values of $\xi^e_S$ and $\xi^q_S$ depend on the
choice of the input value of $\Lambda_S$.
Only the ratio $\xi^e_S/\xi^q_S$ is substantial for the fitting of 
the quark and charged lepton mass.
(However, as we state in the Sec.~\ref{sec:8}, the neutrino mass 
difference between $m_{\nu 3}$ and $m_{\nu 2}$ rapidly varies in
the range III. Therefore, in the neutrino mass matrix, the choice of
the input value $\Lambda_S$ is important.) 
We can obtain 
\begin{equation}
\xi^e_S(\Lambda_X)/\xi^q_S(\Lambda_X) \simeq 1.5 \ ,
\end{equation}
for any initial values of $\xi^e_S(\Lambda_X)$ with $O(1)$.
The values (6.10) are nearly in agreement with the values 
$z_1=0.01622$, $z_2=0.2357$,
and $z_3=0.9717$ which are obtained from Eq.~(1.7)
at $\mu=m_Z$.
We can see that the effect of the evolution is not so large
for $Z^e$.

The value of the parameter $b_d(\Lambda_X)$ is determined from 
the fitting of the observed down-quark mass ratios $m_d/m_s$ 
and $m_s/m_b$ and the CKM matrix parameter $|V_{us}(m_Z)|=0.22$.
In Fig.~\ref{bd-nonSUSY}, 
we illustrate the mass ratios $m_d(\mu)/m_s(\mu)$
and $m_s(\mu)/m_b(\mu)$ and the CKM parameter
$|V_{us}(\mu)|$ at $\mu=m_Z$ versus the parameters $b_d$
and $\beta_d$, where we have re-defined the complex parameter
$b_d$ by $b_d e^{i\beta_d}$ with two real parameters.
For convenience, in Fig.~\ref{bd-nonSUSY}, the quantities
are expressed in the unit of the corresponding observed values
at $\mu=m_Z$ (for example, in Fig.~\ref{bd-nonSUSY},
the curve $m_d/m_s$ denotes $[m_d(\mu)/m_s(\mu)]_{\mu=m_Z}
/[m_d/m_s]_{observed}$).
We obtain
\begin{equation}
b_d(\Lambda_X)=-1.20 \ , \ \ \ 
\beta_d(\Lambda_X)=19.2^\circ \ ,
\end{equation}
which give the following predictions at $\mu=m_Z$:
\begin{eqnarray}
&&m_u(m_Z)=2.60 \times 10^{-3}      \ {\rm GeV} \ , \ \ \nonumber\\
&&m_c(m_Z)=6.92 \times 10^{-1}   \ {\rm GeV}\ , \nonumber\\
&&m_t(m_Z)=182 \ {\rm GeV} \ , \nonumber \\
&&m_d(m_Z)=4.38 \times 10^{-3}      \ {\rm GeV} \ , \nonumber\\
&&m_s(m_Z)=9.84 \times 10^{-2}   \ {\rm GeV}\ , \\
&&m_b(m_Z)=3.02 \ {\rm GeV} \ , \nonumber \\
&&m_e(m_Z)=4.90 \times 10^{-4}      \ {\rm GeV} \ , \nonumber\\
&&m_\mu(m_Z)=1.03 \times 10^{-1}   \ {\rm GeV}\ , \nonumber\\
&&m_\tau(m_Z)=1.76 \ {\rm GeV} \ . \nonumber 
\end{eqnarray}
The experimental values corresponding to the 
results (6.13) are as follows
\cite{qmass}:
\begin{eqnarray}
&& m_u(m_Z)=(2.33^{+0.42}_{-0.45}) \times 10^{-3} \ {\rm GeV} \ , \nonumber\\
&& m_c(m_Z)=(6.85^{+0.56}_{-0.61}) \times 10^{-1}   \ {\rm GeV}\ , \nonumber\\
&& m_t(m_Z)=(181\pm 13) \ {\rm GeV} \ , \nonumber\\ 
&& m_d(m_Z)=(4.69^{+0.60}_{-0.66}) \times 10^{-3}  \ {\rm GeV} \ , \nonumber\\
&& m_s(m_Z)=(0.934^{+0.118}_{-0.130}) \times 10^{-1} \ {\rm GeV}\ , \\
&& m_b(m_Z)=(3.00 \pm 0.11) \ {\rm GeV} \ ,  \nonumber\\
&& m_e(m_Z)=(4.8684727\pm 0.00000014) \times 10^{-4} \ {\rm GeV} \ , 
\nonumber\\
&& m_\mu(m_Z)=(1.0275138 \pm 0.0000033) \times 10^{-1}  \ {\rm GeV}\ ,
\nonumber\\ 
&& m_\tau(m_Z)=(1.7467 \pm 0.0003) \ {\rm GeV} \ .\nonumber 
\end{eqnarray}
The results (6.13) is in agreement with the observed values (6.14) within 
the experimental errors.

The predicted values of $|V_{ij}|$ depends on the phase  parameters
$\delta_i^f$ given by Eq.~(1.4).
Only when we take those as (1.8) (at $\mu=\Lambda_X$), we can obtain
reasonable values of $|V_{ij}|$.
For example, for $\delta_3^u -\delta_3^d=\pi$, we obtain the
predictions at $\mu=m_Z$
\begin{eqnarray}
&& |V_{us}|=0.220 \ , \ \ |V_{cb}|=0.0668  \ , \nonumber\\ 
&& |V_{ub}/V_{cb}|=0.0558 \ , \nonumber\\ 
&& |V_{td}|=0.0177  \ ,  \\
&& J=3.25 \times 10^{-5}    \ . \nonumber
\end{eqnarray}
The observed  values \cite{PDG00} are 
\begin{eqnarray}
&&|V_{us}|=0.2196\pm 0.0023 \ , \ \ \nonumber\\
&&|V_{cb}|=0.0402\pm 0.0019 \ , \ \ \\
&&|V_{ub}/V_{cb}|=0.090 \pm 0.025 \ . \nonumber
\end{eqnarray}
Although the results (6.15) are roughly consistent with experiments, 
the value $|V_{cb}|=0.066$ is somewhat large compared with
the observed value $|V_{cb}|=0.040$.
This discrepancy can be adjusted by considering a small deviation from
$\pi$ of the relative phase $\delta_3^u -\delta_3^d$ as demonstrated  
in Ref.~\cite{KFptp}. 

Related to the phenomenological requirement (1.8),
it is interesting  to consider that $Y_L^u$ which is the coefficient
of the Higgs scalar $\widetilde{\phi}_L$ is related to $Y_L^d$ which
is the  coefficient of the scalar $\phi_L$ as
\begin{equation}
Y_L^u(\Lambda_X) =[Y_L^d(\Lambda_X)]^\dagger \ . 
\end{equation}
Then, the relations (1.8) mean that $(Y_L^f)_{11}$ and $(Y_L^f)_{22}$ 
are  real, while $(Y_L^f)_{33}$ is almost pure imaginary.
We take
\begin{equation}
(Z^u)^\dagger = Z^d ={\rm diag}(z_1, z_2, z_3 e^{i\delta_3}) \ .
\end{equation}
The parameter $\delta_3$ ($=\delta_3^d =-\delta_3^u$) does not 
affect the masses, but only the CKM mixings.
It is interesting to consider that the parameter $\delta_3(\Lambda_X)$
takes its value such as the CKM mixings become minimum, i.e.,
such as the value $\sum_{i\neq j}|V_{ij}(\Lambda_X)|^2$ takes
the minimum.
This requirement gives the initial value $\delta_3(\Lambda_3)=84^\circ$
(see Fig.~\ref{d3-vinjx}).
Then, we obtain the predictions of $|V_{ij}|$ at $\mu=m_Z$
\begin{eqnarray}
&& |V_{us}| = 0.220 \ , \ \ |V_{cb}| =0.0418 \ , \nonumber \\
&& |V_{ub}/V_{cb}|=0.0726 \ ,\nonumber\\ 
&& |V_{td}|=0.0109  \ ,  \\ 
&& J=2.38 \times 10^{-5}    \ . \nonumber
%
\end{eqnarray}
which is in excellent agreement with the experimental values (6.16).
In Fig.~\ref{d3-vij}, we  illustrate the predicted values 
$|V_{ij}(m_Z)|$
versus $\delta_3(\Lambda_X)$.
As seen in  Fig.~\ref{d3-vij}, the value of $\delta_3(\Lambda_X)$ 
at which $\sum_{i\neq j}|V_{ij}(\Lambda_X)|^2$ takes
the minimum also gives the minimum of the CKM mixings at $\mu=m_Z$.

\section{Numerical results in the SUSY model}
\label{sec:7}


The behavior of $\xi_L^u(\mu)$ in the SUSY model is 
somewhat different from that in the non-SUSY model.
Since in the SUSY model, the top quark mass $m_t(\mu)$ is
given by
\begin{equation}
m_t(\mu) = \frac{1}{\sqrt{3}} \xi^u_L(\mu) \frac{v_L}{\sqrt{2}} 
\sin\beta \ ,
\end{equation}
where $v_L/\sqrt{2}= 174$ GeV and $\tan\beta\equiv \tan\beta_L
=v_L^u/v_L^d$, the initial value of $\xi^u_L(m_Z)$
in the SUSY model corresponds to
\begin{equation}
\left[\xi^u_L(m_Z)\right]_{SUSY}=\left[\xi^u_L(m_Z)\right]_{nonSUSY}
\frac{1}{\sin\beta} \ .
\end{equation}
However, this does not mean $[\xi^u_L(\Lambda_X)]_{SUSY}=
[\xi^u_L(\Lambda_X)]_{nonSUSY}/{\sin\beta}$, because the
behavior of  $[\xi^u_L(\mu)]_{SUSY}$ is considerably different 
from that of $[\xi^u_L(\mu)]_{nonSUSY}$.
In Fig.~\ref{xi-SUSY}, we illustrate the behavior of $\xi_L^u(\mu)$
in the SUSY model for the case of $\tan\beta=3.5$.
If we take $\tan\beta < 2.5$, the initial value of $\xi_L^u(m_Z)$
becomes $\xi_L^u(m_Z,\tan\beta>2.5) > \xi_L^u(m_Z,\tan\beta=2.5)$ 
from Eq.~(7.2), so that the curve of $\xi_L^u(\mu)$ will be illustrated
in the upper side of the curve given in Fig.~\ref{xi-SUSY}.
Therefore, for a case with a small value of $\tan\beta$, the
Landau pole of $\xi^u_L(\mu)$ appears at a relatively lower energy scale.
We consider that the model should be calculable perturbatively, 
so that a case with such a large value of $\xi_L^u$ should be ruled out.
As seen in Fig.~\ref{xi-SUSY}, since the model gives, in general, 
$\xi_L^u(\Lambda_X)>\xi_L^u(\mu)$ ($m_Z < \mu < \Lambda_X$), 
the value $\xi_L^u(\Lambda_X)$ should, at least, be 
$[\xi_L^u(\Lambda_X)]^2/4\pi <1$, i.e., 
$\xi_L^u(\Lambda_X) < \sqrt{4\pi} = 3.54$.
However, when we take contributions from the higher order
corrections into consideration, even the value $\xi_L^u(\Lambda_X)=3.0$
is still dangerous.
Therefore, we put the constraint $\xi_L^u(\Lambda_X)=2.0$ for the results
of the present one loop calculation.
In Fig.~\ref{mt-tanb}, we illustrate the predicted value of $m_t(m_Z)$
for the initial values $\xi_L^u(\Lambda_X)=\sqrt{4\pi}=3.54$ and 
$\xi_L^u(\Lambda_X) \alt 2.0$, where we have used the input values
\begin{equation}
\Lambda_X=2\times 10^{16} \ {\rm GeV} \ , \ \ 
\Lambda_S=6\times 10^{13} \ {\rm GeV} \ .
\end{equation}
The value of $\Lambda_S$ has been chosen as the neutrino mass-squared
difference $\Delta m_{32}^2$ is of the order of $(10^{-3}-10^{-2})$ eV$^2$.

{}From Fig.~\ref{mt-tanb}, we conclude that the value of $\tan\beta$ 
must be
\begin{equation}
\tan\beta \agt 3 \ .
\end{equation}

Prior to the numerical investigation of the evolutions in 
the SUSY model, in order to see the difference between the
parameter structures in the non-SUSY and SUSY models, let us 
give a rough sketch for the parameters in the case of the SUSY model 
by neglecting the evolution effects.
The quark mass matrices $M_u$ and $M_d$ are given by
\begin{equation}
\left( M_u \right)_{ij} =\sum_{k=1}^2 
\left(\widetilde{Z}\right)_{ik}
\left(\widetilde{Z}\right)_{jk} \left(O^u\right)_{kk}
\frac{\xi^u_L \xi^u_R}{\xi^u_S}
\frac{\Lambda_L \Lambda_R}{\Lambda_S} \sin\beta \ ,
\end{equation}
\begin{equation}
\left( M_d \right)_{ij} =\sum_{k=1}^3 
\left(\widetilde{Z}\right)_{ik}
\left(\widetilde{Z}\right)_{jk}  \left(O^d\right)_{kk}
\frac{\xi^d_L \xi^d_R}{\xi^d_S}
\frac{\Lambda_L \Lambda_R}{\Lambda_S} \cos\beta \ ,
\end{equation}
where $O^u={\rm diag}(1,1)$,  $O^d={\rm diag}(1,1, 1/(1+3b_d))$, 
and $\widetilde{Z}$ is given by Eq.~(5.14).
Here, for simplicity, we have assumed $\beta_L =\beta_R =\beta_S
\equiv \beta$.
For $\tan\beta > 3$, the factors $\sin\beta$ and $\cos\beta$
are approximated as $\sin\beta\simeq 1$ and 
$\cos\beta \simeq 1/\tan\beta$, respectively.
Obviously, the model with $\xi_S^u =\xi_S^d$ in addition to the
constraint 
\begin{equation}
\xi_{LR}^u (\Lambda_X) = \xi_{LR}^d (\Lambda_X) \equiv
\xi_{LR}^q (\Lambda_X) \ ,
\end{equation}
is ruled out, because we cannot fit the up- and down-quark
masses simultaneously due to the existence of the factor
$\cos\beta$.
Therefore, we must consider a model with $\xi_S^u \neq \xi_S^d$
differently from the constraint (6.6) in the non-SUSY model.
If we consider
\begin{equation}
\xi_S^u \simeq \xi_S^q \sin\beta \ , \ \ \ 
\xi_S^d \simeq \xi_S^q \cos\beta \ ,
\end{equation}
then the model becomes similar to the case of the non-SUSY model,
because
\begin{equation}
\frac{\xi^u_L \xi^u_R}{\xi^u_S}
\frac{\Lambda_L \Lambda_R}{\Lambda_S} \sin\beta \simeq
\frac{\xi^d_L \xi^d_R}{\xi^d_S}
\frac{\Lambda_L \Lambda_R}{\Lambda_S} \cos\beta \ ,
\end{equation}
and we will obtain reasonable fittings for the quark masses
and CKM matrix parameters as well as in the non-SUSY model.
Note that for a large value of $\tan\beta$, the value of
$K^d\equiv \xi_L^d \xi_R^d/\xi_S^d$ becomes large because
$K^d \simeq K^u \tan\beta$ from the relation (7.9), so that
we cannot evaluate the RGE (3.12) perturbatively.
We must take the value of $\tan\beta$ near to the lower
bound given by Eq.~(7.4).

When we take the evolution effects into
consideration, the situation is further complicated.
The evolutions of $\xi_L^f(\mu)$, $\xi_R^f(\mu)$ and
$\xi_S^f(\mu)$ in the SUSY model are quite different from those 
in the non-SUSY model.
We illustrate the behaviors of $m_i^f(\mu)/m_i^f(\Lambda_X)$
which correspond to the behaviors of $[\xi_L^f(\mu)\xi_R^f(\mu)
/\xi_S^f(\mu)]/[\xi_L^f(\Lambda_X)\xi_R^f(\Lambda_X)
/\xi_S^f(\Lambda_X)]$ in the non-SUSY model and those
in the  SUSY model in Figs.~\ref{m-nonSUSY} and \ref{m-SUSY}, 
respectively.
In Fig.~\ref{m-SUSY}, we see that the values $m_u(\mu)$
and $m_c(\mu)$ cause rapid changes in the range III.
In the non-SUSY model, the charged lepton mass ratios are 
almost invariant, i.e., $m_e(\mu)/m_\mu(\mu) \simeq
{\rm constant}$ and $m_\mu(\mu)/m_\tau(\mu)\simeq {\rm constant}$, 
while, in the SUSY model, the mass ratio
$m_\mu(\mu)/m_\tau(\mu)$ shows a considerable change
(although $m_e(\mu)\simeq m_\mu(\mu)$ still holds).

The situation is critical for the input values.
If we adhere to the input value $m_t(m_Z)=181$ GeV, then
it is hard to obtain reasonable values of the other
quark mass values $m_c$, $m_u$, $m_b$, $m_s$ and $m_d$
for any parameter values of $\Lambda_S/\Lambda_R$ and
$b_d$.
However, if we take a slightly lower value of $m_t(m_Z)$,
for example, $m_t(m_Z)=168$ GeV [cf. $[m_t(m_Z)]_{observed}
=181 \pm 13$ GeV], we can find the following 
parameter values
\begin{equation}
\tan\beta = 3.5 \ , \ \ \ \ \Lambda_S/\Lambda_R =38 \ ,
\end{equation}
\begin{equation}
z_1=0.01449 \ , \ \ z_2=0.2117 \ , \ \  z_3=0.9772 \ ,
\end{equation}
\begin{equation}
\xi_{LR}^u (\Lambda_X) = \xi_{LR}^d (\Lambda_X) \equiv
\xi_{LR}^q (\Lambda_X) = 1.3 \ , \ \ 
\xi_{LR}^e (\Lambda_X) = 1.0 \ ,
\end{equation}
\begin{equation}
\xi_{S}^u (\Lambda_X) = 1.7 \ , \ \ 
\xi_{S}^d (\Lambda_X) = 0.50 \ , \ \ 
\xi_{S}^e (\Lambda_X) = 1.0 \ ,
\end{equation}
\begin{equation}
b_d = -1.2 \ , \ \ \ \beta_d = 19.4^\circ \ ,
\end{equation}
which leads to the following quark and charged lepton
masses and CKM matrix parameters:
\begin{eqnarray}
&&m_u(m_Z)=2.47 \times 10^{-3}   \ {\rm GeV} \ , \nonumber \\ 
&&m_c(m_Z)=6.46 \times 10^{-1}   \ {\rm GeV}\ , \nonumber \\ 
&&m_t(m_Z)=167 \ {\rm GeV} \ , \nonumber \\ 
&&m_d(m_Z)=4.49 \times 10^{-3}  \ {\rm GeV} \ , \nonumber \\ 
&&m_s(m_Z)=1.00 \times 10^{-1}   \ {\rm GeV}\ ,  \\ 
&&m_b(m_Z)=2.83 \ {\rm GeV} \ , \nonumber \\ 
&&m_e(m_Z)=4.87 \times 10^{-4}   \ {\rm GeV} \ , \nonumber \\ 
&&m_\mu(m_Z)=1.03 \times 10^{-1}   \ {\rm GeV}\ , \nonumber \\ 
&&m_\tau(m_Z)=1.75 \ {\rm GeV} \ ,\nonumber  
\end{eqnarray}
\begin{eqnarray}
&&|V_{us}|=0.220 \ , \ \ |V_{cb}|=0.0665 \ , \ \ \nonumber\\
&&|V_{ub}/V_{cb}|=0.0603  \ , \nonumber\\
&&|V_{td}|=0.0179      \ , \\
&&J=3.38 \times 10^{-5}   \ .\nonumber
\end{eqnarray}
The values $|V_{ij}|^2$ are again desirably adjustable
by the phase parameter $\delta_3$ defined by (6.18).

\section{Evolution of the Neutrino Mass Matrix}
\label{sec:8}


The evolution of the neutrino mass matrix $M_\nu=K^\nu
\langle \phi_L^0\rangle^2/\langle\Phi^0\rangle$ is
described by the RGE (3.13).
Since the coefficient $H_{KL}^\nu$ in the ranges I and II
is given by $H_{KL}^\nu=\lambda_{HL}$, (4.28), for the 
non-SUSY model, and by $H_{KL}^\nu=0$, (A.15) and (A.24),
for the SUSY model, the form of the matrix $K^\nu$ 
at $\mu=\Lambda_L$ does not
vary from that at $\mu=\Lambda_S$, so that the mass ratios
and mixing matrix $U_\nu$ are also invariant.
Since the coefficients $H_{KL}^e$ and $H_{KR}^e$ in the
charged lepton sector are given by 
$H_{KL}^e=H_{KR}^e=0$ in the ranges I and II for the 
non-SUSY and SUSY models, the form of the charged lepton
mass matrix $M_e$ is also invariant below $\mu=\Lambda_S$.
Therefore, the Maki-Nakagawa-Sakata (MNS) \cite{MNS} matrix 
$U=U_{eL}^\dagger U_\nu$ is invariant in the ranges I and II.
Note that in the conventional model, the neutrino seesaw
mass matrix can vary the from.
The neutrino mass matrix in the present model can vary the 
form only in the range III ($\Lambda_S < \mu \leq \Lambda_X$).
The reason is that in the conventional model the scalar 
$\phi_L^+$ couples to $\overline{\nu}_L e_R$,
while that in the present model couples to 
$\overline{\nu}_L E_R$, so that the contribution of $\phi_L$
to $H_{KL}^\nu$ in the latter case is decoupled below 
$\mu=\Lambda_S$.

For the numerical study, the case with $b_\nu=-1/2$ is most
interesting, because the inverse matrix of 
$Y_S^\nu(\Lambda_X)=\xi_S^\nu(\Lambda_X) [{\bf 1} +
3b_\nu(\Lambda_X) X]$ with $b_\nu(\Lambda_X) =-1/2$ has
the form
\begin{equation}
\left[ Y_S^\nu(\Lambda_X) \right]^{-1} =
-\frac{1}{\xi_S^\nu(\Lambda_X) } 
\left(
\begin{array}{ccc}
0 & 1 & 1 \\
1 & 0 & 1 \\
1 & 1 & 0 \\
\end{array} \right)  \ ,
\end{equation}
so that
\begin{equation}
Y_L^\nu (Y_S^\nu)^{-1}(Y_L^\nu)^T = -
\frac{\xi_L^\nu(\Lambda_X)]^2}{\xi_S^\nu(\Lambda_X)}
\left(
\begin{array}{ccc}
0 & z_1 z_2 & z_1 z_3 \\
z_1 z_2 & 0 & z_2 z_3 \\
z_1 z_3 & z_2 z_3 & 0 \\
\end{array} \right)  \ .
\end{equation}
The form (8.2) is well known as the Zee-type \cite{Zee} mass 
matrix, which can lead to a large mixing \cite{mixing-Zee}.

The mass eigenvalues $m_{\nu i}$ and mixing matrix $U$ at 
$\mu=\Lambda_X$ are given by \cite{nu-Koide}
\begin{eqnarray}
&&m_{\nu 1}\simeq -2 z_1^2 m_0^\nu \ , \nonumber \\ 
&&m_{\nu 2}\simeq - \left[z_2 z_3 -\left( 
1 -\frac{z_3}{2 z_2}\right)
z_1^2 \right] m_0^\nu \ ,  \\  
&&m_{\nu 3} \simeq  \left[z_2 z_3 +\left( 
1+\frac{z_3}{2 z_2} \right)
z_1^2  \right] m_0^\nu \ , \nonumber 
\end{eqnarray}
\begin{equation}
m_0^\nu= \frac{(\xi_L^\nu)^2}{\xi_S^\nu}
\frac{\Lambda_L^2}{\Lambda_S} \ ,
\end{equation}
\begin{equation}
U=\left(
\begin{array}{ccc}
1 & \frac{1}{\sqrt{2}} \frac{z_1}{z_2} (1-z_2) & 
\frac{1}{\sqrt{2}} \frac{z_1}{z_2} (1+z_2) \\
-\frac{z_1}{z_2} & \frac{1}{\sqrt{2}} &
-\frac{1}{\sqrt{2}} \\
-z_1 & -\frac{1}{\sqrt{2}} & \frac{1}{\sqrt{2}} 
\end{array} \right) \ .
\end{equation}
The model with $b_d=-1/2$ gives  highly degenerate 
mass-squared levels $m_{\nu 2}^2 \simeq m_{\nu 3}^2$ and a large 
mixing between $\nu_\mu$ and $\nu_\tau$ at $\mu=\Lambda_X$.
Therefore, the model has a possibility that it can give a 
reasonable explanation for the atmospheric neutrino data 
\cite{nu-atm}.

In Figs.~\ref{dm-nonSUSY} and \ref{dm-SUSY}, we illustrate
the behaviors of the mass-squared differences
$\Delta m_{ij}^2=m_i^2 -m_j^2$ in the non-SUSY and SUSY models,
respectively.
As seen in Figs.~\ref{dm-nonSUSY} and \ref{dm-SUSY},
the mass-squared difference $\Delta m_{32}^2$ rapidly increase
according as the energy scale decreases in the range III.
The numerical results are given in Table \ref{T-dm}.
We can see that the neutrino mass ratios are invariant
in the ranges I and II.

As we stated already, the values $(z_1, z_2, z_3)$ 
(therefore, the mass ratios $m_e/m_\mu$ and $m_\mu/m_\tau$)
are almost invariant in the range III, while the ratio
$\Delta m^2_{32}/\Delta m^2_{21}$ is rapidly vary in
the range III.
Although the relations (8.3) give $\Delta m^2_{32} \simeq
4 z_1^2 z_2 z_3 (m_0^\nu)^2$ and  
$\Delta m^2_{21} \simeq (z_2 z_3^2)^2 (m_0^\nu)^2$,
the rapid decrease in the ratio 
$\Delta m^2_{32}/\Delta m^2_{21}$ does not mean the
rapid decrease in the ratio $z_1^2/z_2 z_3$.
The rapid decrease comes from the slight deviation of
the parameter $b_\nu(\mu)$ from
the value $b_\nu (\Lambda_X)=-1/2$.
The value of $b_f(\mu)$ is not invariant in the range
III, although the form of $Y_S^f(\mu)$, 
``the unit matrix plus a democratic matrix", is
invariant.
When we denote
\begin{equation}
b_\nu(\mu) = -\frac{1}{2} \left(1+\varepsilon_\nu(\mu)
\right) \ ,
\end{equation}
the expression (8.1) is replaced with
\begin{equation}
\left[ Y_S^\nu(\mu) \right]^{-1} \simeq 
-\frac{1}{\xi_S^\nu(\mu) } 
\left(
\begin{array}{ccc}
-2 \varepsilon_\nu & 1 & 1 \\
1 & -2 \varepsilon_\nu & 1 \\
1 & 1 & -2 \varepsilon_\nu \\
\end{array} \right)  \ ,
\end{equation}
so that
\begin{equation}
Y_L^\nu (Y_S^\nu)^{-1}(Y_L^\nu)^T \simeq  -
\frac{[\xi_L^\nu(\mu)]^2}{\xi_S^\nu(\mu)}
\left(
\begin{array}{ccc}
-2 \varepsilon_\nu z_1^2 & z_1 z_2 & z_1 z_3 \\
z_1 z_2 & -2 \varepsilon_\nu z_2^2 & z_2 z_3 \\
z_1 z_3 & z_2 z_3 & -2 \varepsilon_\nu z_3^2 \\
\end{array} \right)  \ .
\end{equation}
Therefore, the mass eigenvalues in the range III are
given by
\begin{eqnarray}
&&m_{\nu 1}\simeq -2 (1+3 \varepsilon_\nu ) z_1^2 m_0^\nu 
\ , \nonumber \\ 
&&m_{\nu 2}\simeq - \left[z_2 z_3 -\left( 1-\frac{z_3}{2 z_2}\right)
z_1^2 +\varepsilon_\nu \right] m_0^\nu \ ,  \\  
&&m_{\nu 3} \simeq  \left[z_2 z_3 +\left( 1+\frac{z_3}{2 z_2} \right)
z_1^2 -\varepsilon_\nu  \right] m_0^\nu \ , \nonumber 
\end{eqnarray}
instead of (8.3), and the mass squared differences 
$\Delta m^2_{21}$ and $\Delta m^2_{32}$ are given by
\begin{eqnarray}
&&\Delta m^2_{21} \simeq (z_2 z_3)^2 (m_0^\nu)^2 \ ,
\nonumber \\
&&\Delta m^2_{32} \simeq 4 z_2 z_3 (z_1^2 -\varepsilon_\nu)
(m_0^\nu)^2 \ .
\end{eqnarray}
Note that the approximate expression (8.19) tell us that
$\Delta m^2_{32}(\mu)$ takes a zero between $\mu=\Lambda_X$
and $\mu=\Lambda_S$ because $\varepsilon_\nu(\Lambda_X)=0
< z_1^2(\Lambda_X) \simeq z_1^2(\Lambda_S) < 
\varepsilon_\nu(\Lambda_S)$, e.g., 
$\varepsilon_\nu(\Lambda_S)= 7.3\times 10^{-2}$ and
$z_1^2(\Lambda_X) \simeq z_1^2(\Lambda_S) \simeq 2.6 \times
10^{-4}$ for the non-SUSY and 
$\varepsilon_\nu(\Lambda_S)= 1.1\times 10^{-2}$,
$z_1^2(\Lambda_X) \simeq 2.1 \times 10^{-4}$ and 
$z_1^2(\Lambda_S) \simeq 2.6 \times 10^{-4}$ for the SUSY 
model.
In fact, we can see this at a point which is very close to
$\mu=\Lambda_X$ in Figs.~\ref{dm-nonSUSY} 
and \ref{dm-SUSY}.
Thus, the value of $\Delta m^2_{32}(\mu)$ is highly sensitive
to the value of $\varepsilon_\nu(\mu)$, although 
$\Delta m^2_{21}(\mu)$ is not so.

In general, since the mixing angle $\theta_{23}$ is 
given by
\begin{equation}
\sin 2\theta_{23} \simeq \frac{2 (M_\nu)_{23}}{
m_{\nu 3}-m_{\nu 2}} \ ,
\end{equation}
the mixing angle $\theta_{23}$ in the conventional
democratic type neutrino mass matrix model is
sensitive to the energy scale \cite{dem-nu-evol}, because 
$\Delta m_{32}^2(\mu)$ has a large energy scale
dependency.
In contrast to the conventional model, the mixing angle
$\theta_{23}$ in the present model does not so drastically
vary.
The reason is as follows: the neutrino mass matrix $M_\nu$
in the present ``democratic" seesaw model is not 
democratic, i.e., the form of $M_\nu$ is given by Eq.~(8.2).
In fact, the present model gives not $m_{\nu 2}\simeq m_{\nu 3}$,
but $m_{\nu 2}\simeq -m_{\nu 3}$, so that the evolution effect
on $U_{23}$ is not so sensitive as seen in Eq.~(8.11).

As seen in Table \ref{T-dm}, the model can fit the value
$\Delta m^2_{32}$ to the atmospheric neutrino data
\cite{nu-atm} $(\Delta m_{32}^2)_{observ}=3.2
\times 10^{-3}$ eV$^2$ by adjusting the value of $\Lambda_S$,
but it cannot give any explanation of the solar
neutrino data \cite{nu-solar}, because of 
$\Delta m_{21}^2 \gg \Delta m_{32}^2 \equiv \Delta m_{atm}^2$.
We must introduce a further mechanism for the explanation of
the solar neutrino data, for example, as discussed in
Ref.~\cite{nu-DUSM}.
However, since the purpose of the present model is
not to propose a plausible neutrino mass matrix model
in the framework of the USM, but to see the 
characteristic features of the neutrino mass matrix
evolution in contrast to the conventional seesaw model.
Therefore, we do not touch the numerical
fitting furthermore.

\section{Conclusions}
\label{sec:9}


In conclusion, we have investigated the evolutions of
the quark and lepton mass matrices $M_f$ ($f=u,d,\nu,e$)
in the universal seesaw model with ${\rm det}M_F=0$ 
in the up-quark sector $F=U$.

The assumptions which have made in the present paper
are  classified into the following three categories:

\noindent
(A) Basic assumptions in the universal seesaw model
with ${\rm det}M_U=0$; 

\noindent
(B) Basic assumptions in the democratic seesaw model
 \cite{KFzp,KFptp}
(we have taken the model as a more concrete one of 
the universal seesaw model with ${\rm det}M_U=0$ in order to give
an explicit evaluation of the universal seesaw model);

\noindent
(C) Tentative assumptions for convenience of the numerical 
evaluation.

The assumptions in the category (A) are as follows:

\noindent
(A1) ${\rm SU(3)}_c \times{\rm SU(2)}_L \times 
{\rm SU(2)}_R \times {\rm U(1)}_Y \times {\rm U(1)}_X$
gauge symmetries with the symmetry breaking pattern (2.1);

\noindent
(A2) Hypothetical heavy fermions $F=(U,D,N,E)$ which belong
to $(1,1)$ of ${\rm SU(2)}_L \times {\rm SU(2)}_R$ and 
acquire masses of the order of $\Lambda_S$ at the energy
scale $\mu=\Lambda_S$ except for $U_{3L}$ and $U_{3L}$.

In the present model, therefore, the quark and charged 
lepton mass matrices $M_f$ ($f=u,d,e$) and neutrino mass 
matrix $M_\nu$ are given by
\begin{equation}
M_f=Y_L^f (Y_S^f)^{-1}(Y_R^f)^\dagger (\Lambda_L\Lambda_R/\Lambda_S)
\ ,
\end{equation}
\begin{equation}
M_\nu = Y_L^\nu (Y_S^\nu)^{-1}(Y_L^\nu)^T (\Lambda_L^2/\Lambda_S)
\ ,
\end{equation}
except for the top quark, where $\Lambda_L=\langle\phi^0_L\rangle$,
$\Lambda_R=\langle\phi^0_R\rangle$ and $\Lambda_S=\langle\Phi^0\rangle$.
The evolutions below $\mu=\Lambda_S$ are described by the RGE (3.12)
and (3.13) for the seesaw operators.
On the other hand, the top quark mass $m_t(\mu)$ given by the
expression (3.18) is still described by RGE (3.6) for the Yukawa 
coupling constants below $\mu=\Lambda_S$.
Although the heavy fermions $F$ do not contribute to the evolutions
below $\mu=\Lambda_S$, the third family ``would-be" heavy up-quark
$U_3$ can contribute to the RGE even below $\mu=\Lambda_S$.
However, as far as the $H_{KL}^f$ ($F=\nu, e$) and $H_{KR}^e$ terms
in the lepton sectors are concerned, the would-be heavy quark $U_3$
cannot contribute to those, so that the forms of the mass matrices
$M_\nu(\mu)$ and $M_e(\mu)$ are invariant below $\mu=\Lambda_S$.

The assumptions in the category (B) are as follows:

\noindent
(B1) At a unification scale $\mu=\Lambda_X$, the Yukawa coupling
constants $Y_L^f$ and $Y_R^f$ have the same form, i.e.,
$Y_L^f(\Lambda_X) = Y_R^f(\Lambda_X) \equiv Y_{LR}^f(\Lambda_X)$.

\noindent
(B2) At $\mu=\Lambda_X$, the heavy fermion mass matrices $M_F$ 
(therefore, the Yukawa coupling constants $Y_S^f$) 
[and also the Majorana masses $M_L$ ($M_R$) of the neutral 
fermions $N_L$ ($N_R$)] take a simple diagonal form  (5.1),
the form `` the unit matrix pulse a democratic 
matrix", on the basis on which the Yukawa coupling constants 
$Y_{LR}^f(\Lambda_X)$ are diagonal.
Then, the form (5.1) is invariant under the evolution in the
range III.

\noindent
(B3) The values of the parameter $b_f$ in the matrix $Y_S^f$
given by Eq.~(5.1) are given by $b_e=0$, $b_\nu=-1/2$ and 
$b_u=-1/3$ at $\mu=\Lambda_X$. (The value $b_d$ is kept as
a free parameter in order to fit the up- and down-quark 
masses and CKM matrix parameters reasonably.)

\begin{figure}
\begin{center}
\includegraphics[width=8.6cm]{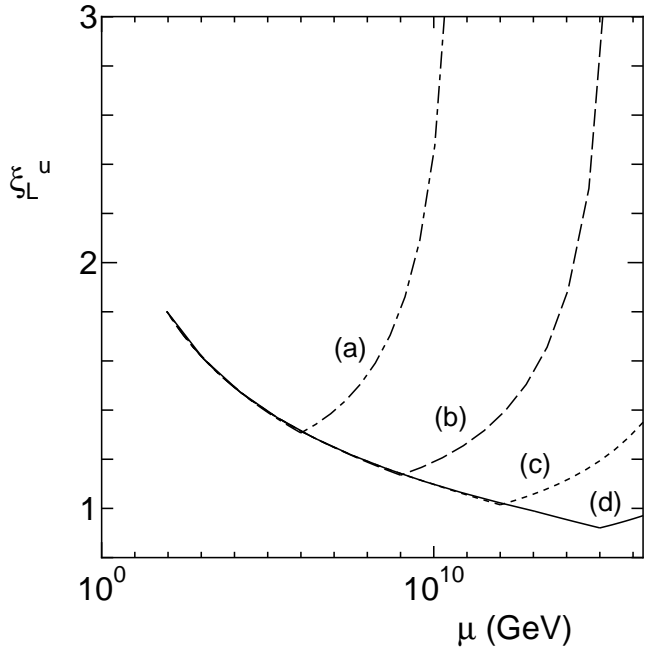}
\end{center}
\caption{
Behaviors of $\xi^u(\mu)$ in a non-SUSY model 
for the cases (a) $\Lambda_S=10^6$ GeV, 
(b) $\Lambda_S=10^9$ GeV, (c) $\Lambda_S=10^{12}$ GeV, 
and (d) $\Lambda_S=10^{15}$ GeV.
The input values are $m_t(m_Z)=181$ GeV and 
$\Lambda_S/\Lambda_R=107$.
}
\label{xi-nonSUSY}

\end{figure}

\begin{figure}
\begin{center}
\includegraphics[width=8.6cm]{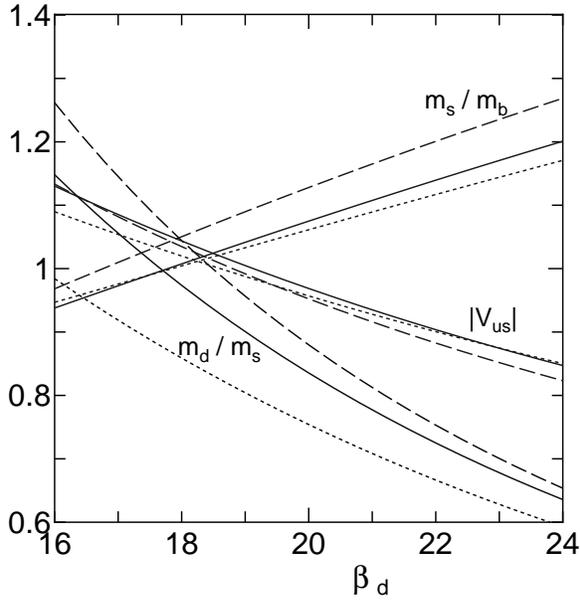}
\end{center}
\caption{
Predictions of $m_d/m_s$, $m_s/m_b$ and $|V_{us}|$, 
and their dependency on the parameters $b_d$ and $\beta_d$.
Here, the mass ratios are denoted in the unit of the
corresponding observed values which are quoted from
Ref.~[10].
The dashed, solid and dotted lines denote $b_d=-1.1$,
$-1.2$ and $-1.3$, respectively.
}
\label{bd-nonSUSY}

\end{figure}

\begin{figure}
\begin{center}
\includegraphics[width=8.6cm]{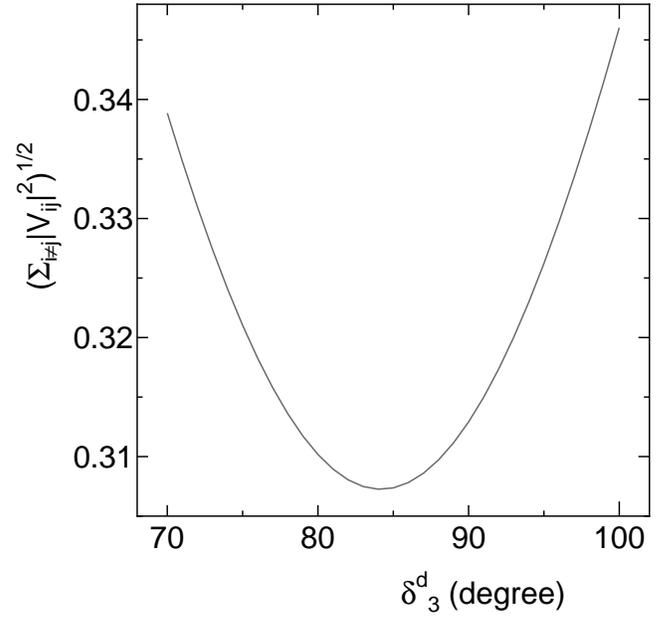}
\end{center}
\caption{
$\sum_{i\neq j} |V_{ij}(\Lambda_X)|^2$ versus
$\delta_3^d(\Lambda_X)$.
}
\label{d3-vinjx}

\end{figure}

\begin{figure}
\begin{center}
\includegraphics[width=8.6cm]{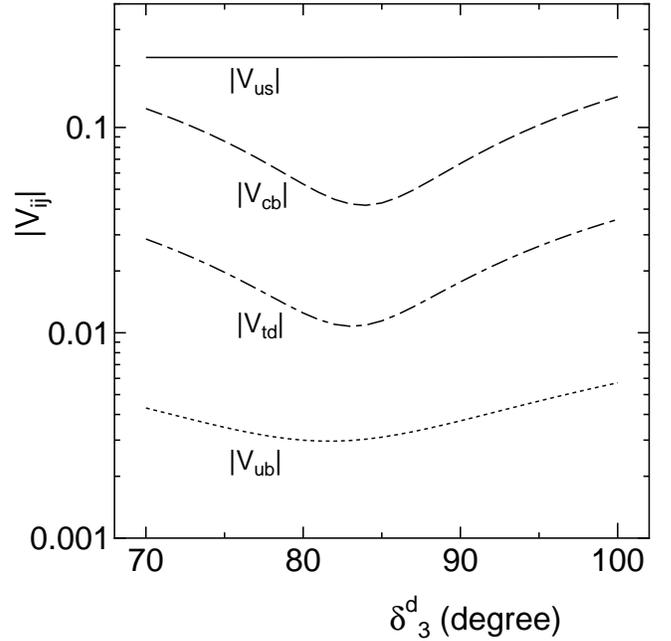}
\end{center}
\caption{
Predicted values of the CKM matrix parameters $|V_{ij}(m_Z)|$ versus
the parameter $\delta_3^d(\Lambda_X)$.
Other input values of the parameters are $\Lambda_S=3\times 10^{13}$
GeV, $\Lambda_S/\Lambda_R=107$, $b_d= -1.2$ and $\beta_d=19.2^\circ$.
}
\label{d3-vij}

\end{figure}

\begin{figure}
\begin{center}
\includegraphics[width=8.6cm]{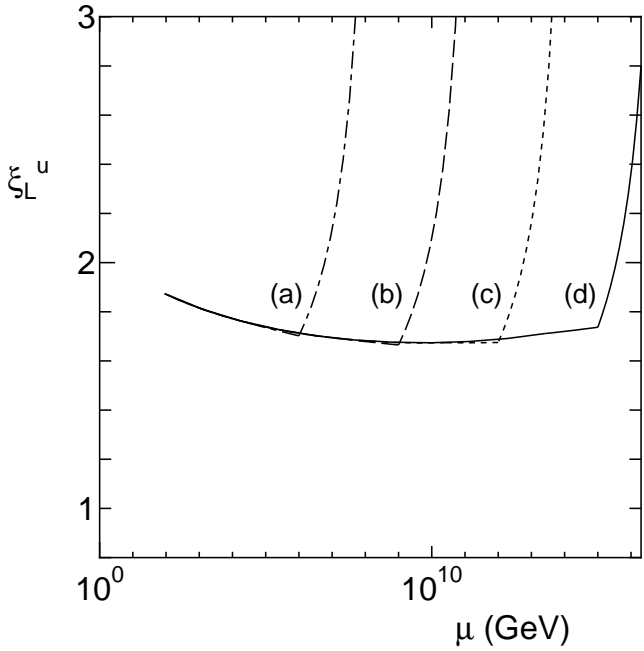}
\end{center}
\caption{
Behaviors of $\xi^u(\mu)$ in a SUSY model 
for the cases (a) $\Lambda_S=10^6$ GeV, 
(b) $\Lambda_S=10^9$ GeV, (c) $\Lambda_S=10^{12}$ GeV, 
and (d) $\Lambda_S=10^{15}$ GeV.
The input values are $m_t(m_Z)=181$ GeV, 
$\Lambda_S/\Lambda_R=38$ and $\tan\beta=3.5$.
}
\label{xi-SUSY}

\end{figure}

\begin{figure}
\begin{center}
\includegraphics[width=8.6cm]{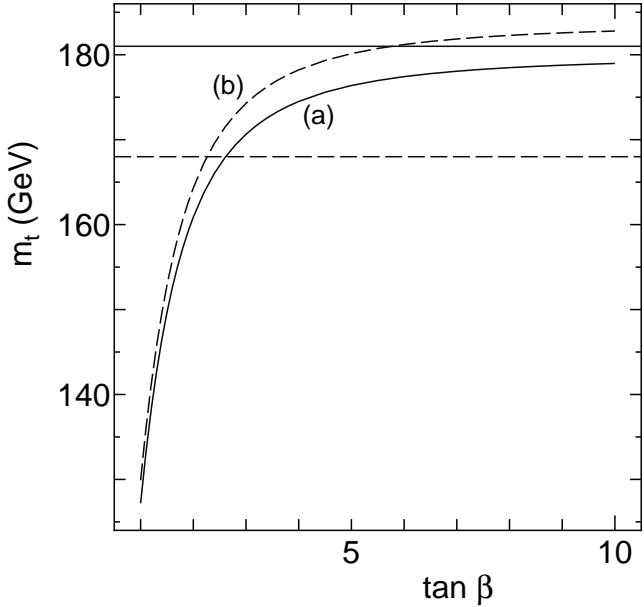}
\end{center}
\caption{
The top-quark mass $m_t(m_Z)$ versus $\tan\beta$ in a SUSY model. 
The solid and broken lines denote the cases with the initial
conditions (a) $\xi_L^u(\Lambda_X)=2.0$ and (b) 
$\xi_L^u(\Lambda_X)=\sqrt{4\pi}=3.54$, respectively.
The other input values are $\Lambda_S=6 \times 10^{13}$ GeV and 
$\Lambda_S/\Lambda_R=38$.
The horizontal solid and broken lines denote the center and lower
values of the observed top quark mass at $\mu=m_Z$, respectively.
}
\label{mt-tanb}

\end{figure}

In this model, the top quark mass $m_t(\mu)$ is given by (5.15).
The behavior of $m_t(\mu)$, i.e., $\xi_L^u(\mu)$,  is given in 
Figs.~\ref{xi-nonSUSY} and \ref{xi-SUSY} for the non-SUSY and
SUSY models, respectively.
We can obtain the constraint on the values of the intermediate
energy scales $\Lambda_R$ and $\Lambda_S$ by considering that
the model should be calculable perturbatively.
In the non-SUSY model, since $\Lambda_S/\Lambda_R \sim 10^2$
from the ratio $m_t/m_c$, we find the constraint
\begin{equation}
10^{10}\ {\rm GeV} < \Lambda_S < 10^{19}\ {\rm GeV} \ ,
\end{equation}
for $\Lambda_X \sim 10^{16}$ GeV.
In the SUSY model, the results highly depend on the input parameter
$\tan\beta$.
{}From the numerical study, we obtain the constraints
\begin{equation}
3 < \tan\beta < 4 \ ,
\end{equation}
\begin{equation}
10^{10}\ {\rm GeV} < \Lambda_S < 10^{19}\ {\rm GeV} \ ,
\end{equation}
for $\Lambda_X \sim 10^{16}$ GeV.
(The above numerical results are slightly dependent
on the assumptions stated below, but the dependence
is not so large.)

The assumptions in the category (C) are as follows:

\noindent
(C1) For convenience of the numerical evaluation,
 $g_{2L}(\mu)=g_{2R}(\mu)$ has been assumed.
Then, we can assert $Y_L^f(\mu) = Y_R^f(\mu)$ 
in the range III ($\Lambda_S < \mu \leq \Lambda_X$) 
as we have shown in Eq.~(4.8).

\noindent
(C2) For evaluation of the non-SUSY model, the initial
condition
\begin{equation}
\xi_{LR}^u(\Lambda_X) = \xi_{LR}^d(\Lambda_X) =
\xi_{LR}^e(\Lambda_X) = \xi_{LR}^\nu(\Lambda_X)
\end{equation}
has been assumed together with the initial condition
(3.5), i.e.,
\begin{equation}
Z_{L/R}^u = Z_{L/R}^d = Z_{L/R}^e = Z_{L/R}^\nu \ .
\end{equation}
However, since there is no solution of
the parameter values for the SUSY model under such a
constraint (9.6), the constraint corresponding to
(9.6) in the SUSY model has been loosened as 
\begin{equation}
\xi_{LR}^u(\Lambda_X) = \xi_{LR}^d(\Lambda_X) \neq
\xi_{LR}^e(\Lambda_X) = \xi_{LR}^\nu(\Lambda_X) \ ,
\end{equation}
although the initial condition (9.7) has still been required
in the SUSY model.

\noindent
(C3) For the non-SUSY model, we have assumed 
$\xi_S^u(\Lambda_X) = \xi_S^d(\Lambda_X) \neq
\xi_S^e(\Lambda_X) = \xi_S^\nu(\Lambda_X)$, but,
for the SUSY model, we have assumed that each value
of $\xi_S^f(\Lambda)$ may be different among them,
because the previous condition is too strong for 
the SUSY model and the up$\leftrightarrow$down symmetry 
is already broken due to the factor $\tan\beta \neq 1$ 
in the SUSY model.

\begin{figure}
\begin{center}
\includegraphics[width=8.6cm]{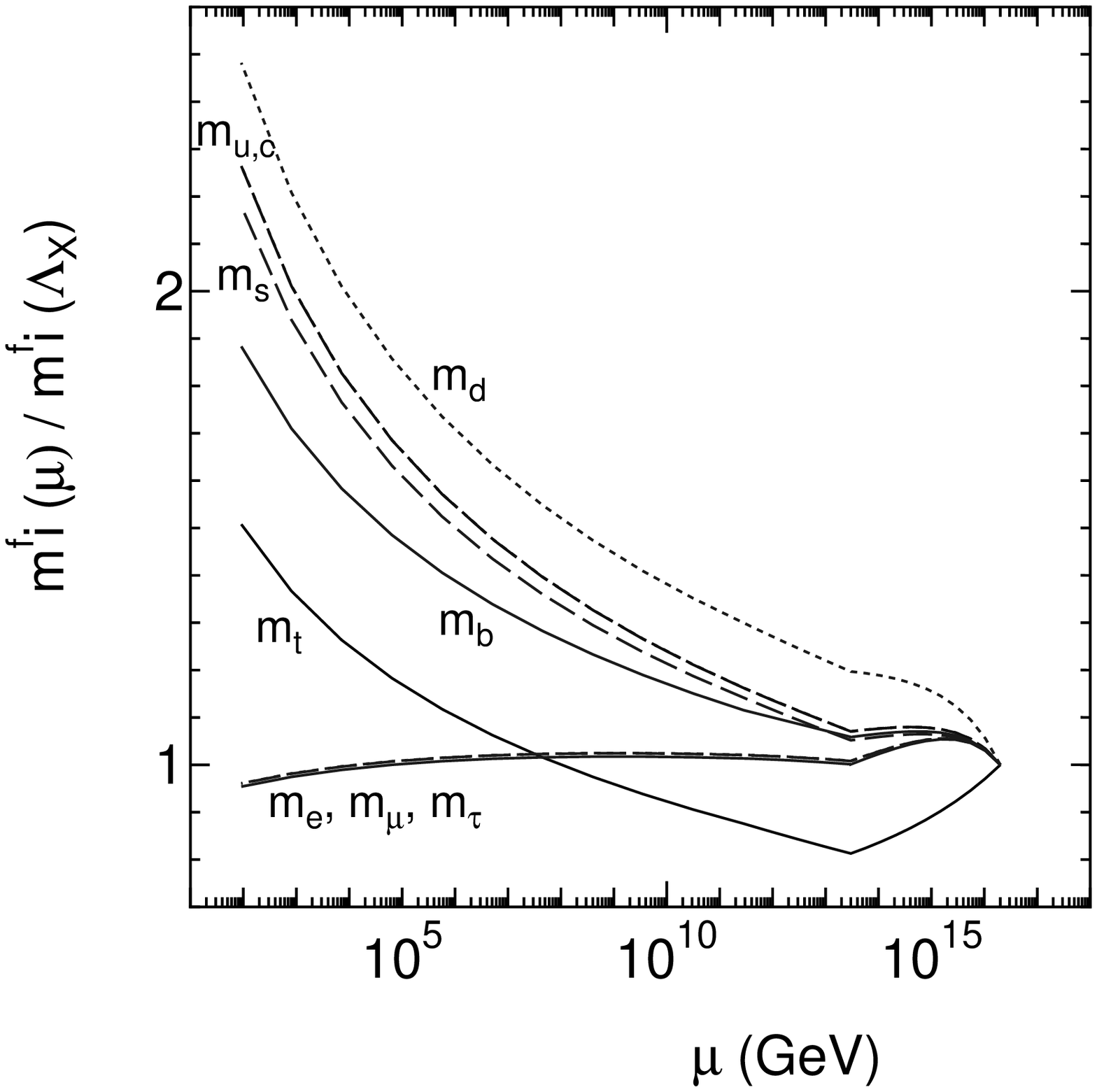}
\end{center}
\caption{
Behaviors of $m^f_i(\mu)/m^f_i(\Lambda_X)$ ($f=u,d,\nu,e$; 
$i=1,2,3$) in the non-SUSY model. 
The dotted, broken and solid lines denote the first,
second and third fermion masses, respectively.
The input parameter values are $\Lambda_S=3\times 10^{13}$ GeV,
$\Lambda_S/\Lambda_R=107$
and $b_d(\Lambda_X)=-1.2 e^{i 19.2^\circ}$.
}
\label{m-nonSUSY}

\end{figure}

\begin{figure}
\begin{center}
\includegraphics[width=8.6cm]{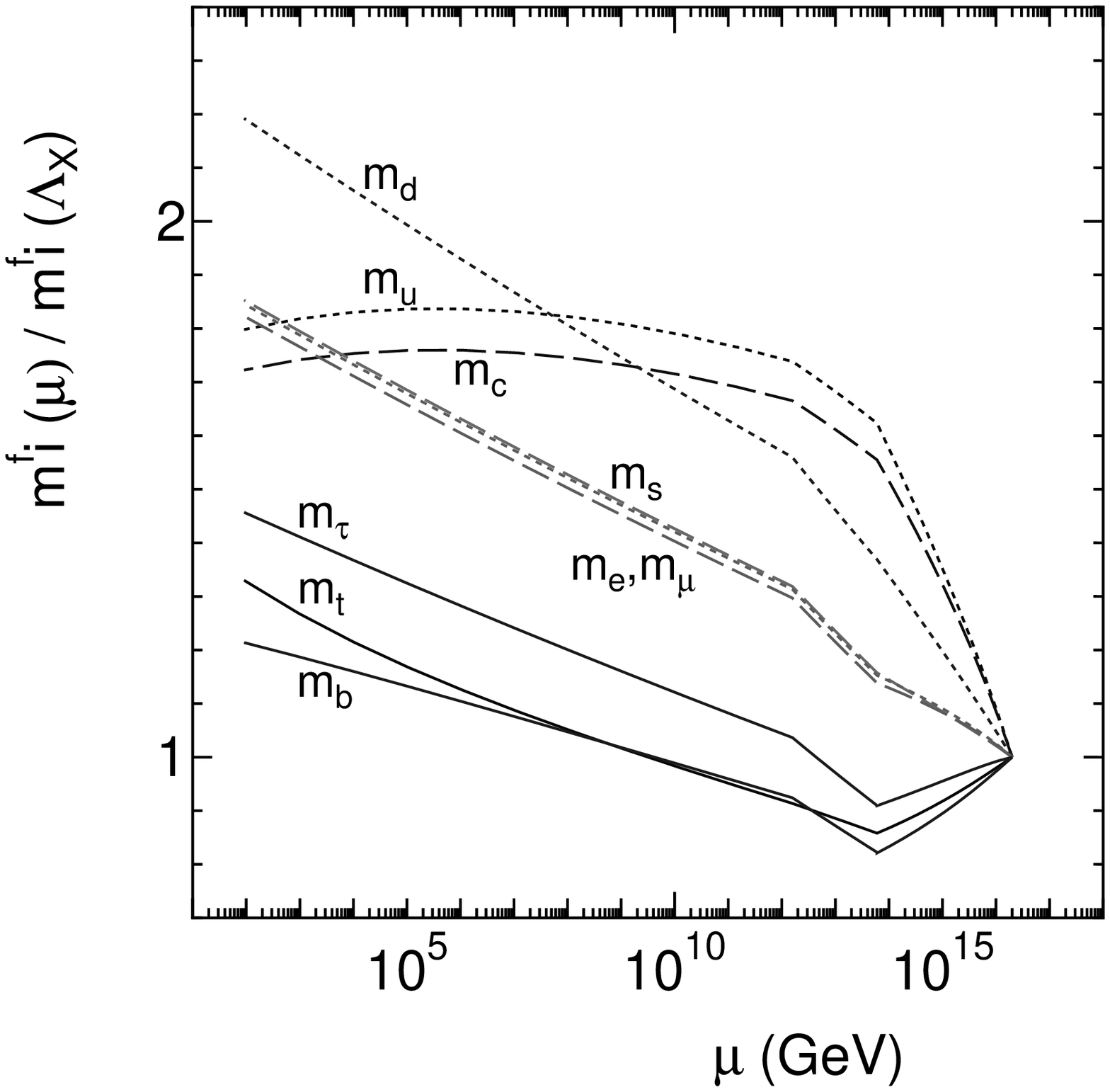}
\end{center}
\caption{
Behaviors of $m^f_i(\mu)/m^f_i(\Lambda_X)$ ($f=u,d,\nu,e$; 
$i=1,2,3$) in the SUSY model. 
The dotted, broken and solid lines denote the first,
second and third fermion masses, respectively.
The input parameter values are $\Lambda_S=6\times 10^{13}$ GeV,
$\Lambda_S/\Lambda_R=38$,
$\tan\beta=3.5$ and $b_d(\Lambda_X)=-1.2 e^{i 19.4^\circ}$.
}
\label{m-SUSY}

\end{figure}

\begin{figure}
\begin{center}
\includegraphics[width=8.6cm]{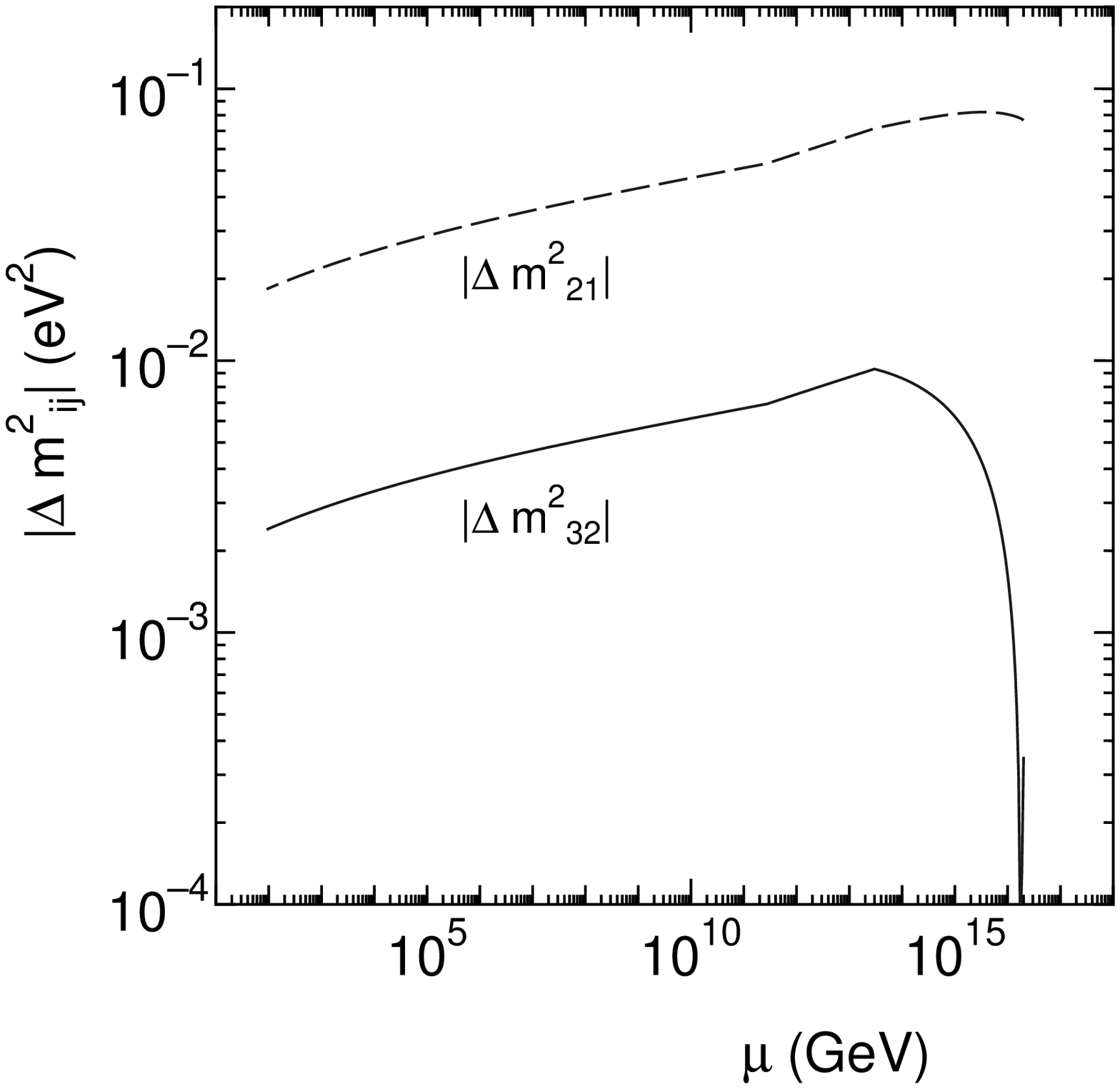}
\end{center}
\caption{
Behavior of $|\Delta m_{ij}^2(\mu)|$ in the non-SUSY model.
The input parameter values are the same as in 
Fig.~7 
with $\xi_A^\nu=\xi_A^e$ ($A=L, R, S$).
}
\label{dm-nonSUSY}

\end{figure}

\begin{figure}
\begin{center}
\includegraphics[width=8.6cm]{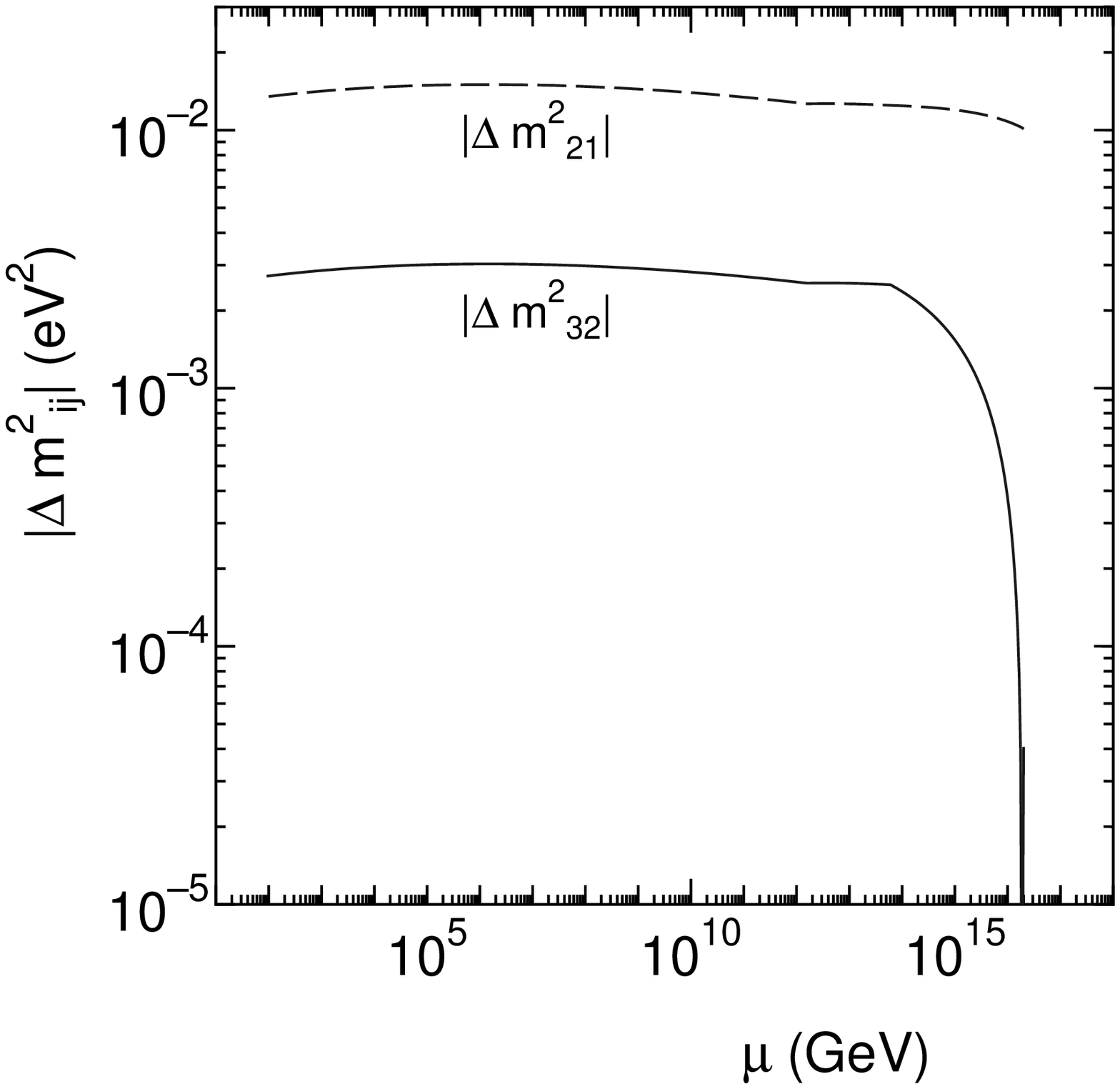}
\end{center}
\caption{
Behavior of $|\Delta m_{ij}^2(\mu)|$ in the SUSY model.
The input parameter values are the same as in 
Fig.~8 
with $\xi_A^\nu=\xi_A^e$ ($A=L, R, S$).
}
\label{dm-SUSY}

\end{figure}

In the conventional model for quark and charged lepton
masses (i.e., not seesaw model), the following 
approximate relations are satisfied in the non-SUSY and 
SUSY models:
\begin{equation}
\frac{(m_u/m_c)_L}{(m_u/m_c)_X} \simeq 
\frac{(m_d/m_s)_L}{(m_d/m_s)_X} \simeq 
\frac{(m_e/m_\mu)_L}{(m_e/m_\mu)_X} \simeq 
\frac{(m_\mu/m_\tau)_L}{(m_\mu/m_\tau)_X} \simeq 1 \ ,
\end{equation}
\begin{equation}
\frac{(m_u/m_t)_L}{(m_u/m_t)_X} \simeq 
\frac{(m_c/m_t)_L}{(m_c/m_t)_X} \simeq 1+ \varepsilon_u
 \ ,
\end{equation}
\begin{equation}
\frac{(m_d/m_b)_L}{(m_d/m_b)_X} \simeq 
\frac{(m_s/m_b)_L}{(m_s/m_b)_X} \simeq 1+ \varepsilon_d
 \ ,
\end{equation}
\begin{equation}
\frac{|V_{cb}(\Lambda_L)|}{|V_{cb}(\Lambda_X)|} \simeq
\frac{|V_{ub}(\Lambda_L)|}{|V_{ub}(\Lambda_X)|} \simeq
\frac{|V_{td}(\Lambda_L)|}{|V_{td}(\Lambda_X)|} \simeq
1 + \varepsilon_d \ ,
\end{equation}
where $(m_u/m_c)_L$ denotes $m_u(\Lambda_L)/m_c(\Lambda_L)$,
and so on.
The relations (9.9)-(9.12) are due to that the Yukawa
coupling constant $y_t$ of the top quark in the
conventional model is very large
compared with the other Yukawa coupling constants.
In the present model, 
as seen in Figs.~\ref{m-nonSUSY} and \ref{m-SUSY}, 
the relations (9.9)-(9.12) are also satisfied
in the range I ($\Lambda_L < \mu \leq \Lambda_R$)
(so that we read the relations (9.9)-(9.12) as 
$X \rightarrow R$).
The values of $\varepsilon_u$ and $\varepsilon_d$ are
approximately given by $\varepsilon_u \sim \varepsilon_d$
for the non-SUSY model, and by $\varepsilon_u \simeq 
-3 \varepsilon_d$ for the SUSY model.
In the range II ($\Lambda_R < \mu \leq \Lambda_S$),
the relations (9.9)-(9.12) are slightly broken.
In the SUSY model, the values show not 
$\varepsilon_u \simeq -3 \varepsilon_d$, but 
$\varepsilon_u \sim \varepsilon_d$ in the range II.
However, in the model with $\Lambda_L/\Lambda_R \gg 1$,
which is required in order to make the neutrino masses
tiny, the evolution effects in the range II are not so
large, so that we can regard that the relations 
(9.9)-(9.12) are still satisfied in the range 
$\Lambda_L < \mu \leq \Lambda_S$, i.e.,
\begin{equation}
D_u(\Lambda_L) \simeq \frac{m_t(\Lambda_L)}{m_t(\Lambda_S)}
(1+\varepsilon_u)({\bf 1} - \varepsilon_u S) D_u(\Lambda_S) 
\ ,
\end{equation}
\begin{equation}
D_d(\Lambda_L) \simeq \frac{m_b(\Lambda_L)}{m_b(\Lambda_S)}
(1+\varepsilon_d)({\bf 1} - \varepsilon_d S) D_d(\Lambda_S) 
\ ,
\end{equation}
\begin{equation}
V(\Lambda_L) \simeq ({\bf 1} + \varepsilon_V S) V(\Lambda_S) 
({\bf 1} + \varepsilon_V S) - 2 \varepsilon_V V_{td} S
\ ,
\end{equation}
where $D_u= {\rm diag}(m_u, m_c, m_t)$, 
$D_d= {\rm diag}(m_d, m_s, m_b)$, and $S$ is defined by
Eq.~(4.9).
In the present model, the value of $\varepsilon_V$ is not
always given by $\varepsilon_V \simeq \varepsilon_d$
because of the presence of the range II.

Also in the ranges I and II,  
differently from the conventional seesaw model 
(for example, see Ref.~\cite{nu-evol}), 
the neutrino mass ratios and mixing angles are not 
affected by the evolution effects:
\begin{equation}
\frac{m_{\nu i}(\Lambda_L)/m_{\nu j}(\Lambda_L) }{
m_{\nu i}(\Lambda_S)/m_{\nu j}(\Lambda_S)}
\simeq 1 \ ,
\end{equation}
\begin{equation}
\frac{V_{ij}(\Lambda_L)}{V_{ij}(\Lambda_S)}
\simeq 1 \ .
\end{equation}
Note that the relation (9.16) does not mean
$\Delta m_{ij}^2(\Lambda_L)/\Delta m_{ij}^2(\Lambda_S)
\simeq 1$.
However, the ratio $\Delta m_{21}^2/\Delta m_{32}^2$
is again invariant in the ranges I and II.

In the range III ($\Lambda_S < \mu \leq \Lambda_X$),
the relations (9.9)-(9.12) [(9.13)-(9.15)] and
(9.16)-(9.17) are not satisfied at all.
For example, the behavior of $\Delta m^2_{32}(\mu)$
is highly sensitive to the value $\varepsilon_\nu(\mu)$
and is given by Eq.~(8.10).
In other words, the differences of the numerical
behaviors of the quark masses, CKM matrix parameters
and neutrino mass squared differences 
from those in the conventional model are 
substantially formed in the range III.

Note that the mass ratios $m_e/m_\mu$ and 
$m_u/m_c$ are almost constant (although the ratio $m_u/m_c$ is
slightly changed in the SUSY model), so that the phenomenologically
well-satisfied relation (1.14) still holds under the evolutions.

For the neutrino mass matrix $M_\nu$, we have investigate the model
with $b_\nu (\Lambda_X)=-1/2$, which leads to a large mixing
$\sin^2 \theta_{23}\simeq 1$.
Although the mass-squared difference $\Delta m_{32}^2(\mu)$ is 
highly sensitive to the energy scale $\mu$ in the range III 
($\Lambda_S <\mu \leq \Lambda_X$), the mixing angle $\theta_{23}$
is not sensitive to the energy scale.
In contrast to the conventional seesaw neutrino mass matrix, 
note that the present neutrino mass matrix $M_\nu$ is form-invariant 
below $\mu=\Lambda_S$, so that the neutrino mass ratios and 
mixings are invariant below $\mu=\Lambda_S$.

In the present paper, we have assumed ${\rm SU(3)}_c\times
{\rm SU(2)}_L \times {\rm SU(2)}_R \times {\rm U(1)}_{LR} \times 
{\rm U(1)}_X$ symmetries above $\mu=\Lambda_S$. 
As seen in Figs.~\ref{xi-nonSUSY} and \ref{xi-SUSY}, in general,
the rapid increasing of the Yukawa coupling constant $Y_L^u(\mu)$ 
causes above $\mu=\Lambda_S$, although we have been able to find 
a set of the reasonable parameter values without having the Landau
pole below $\mu=\Lambda_X$.
The rapid increasing is mainly due to the rapid increasing of 
the gauge coupling constant $g_1$ above $\mu=\Lambda_S$.
If we want to build a unification model with a unified gauge
symmetry G, we may consider that the U(1) symmetry is embedded
into the unified symmetry G.
(For example, see an ${\rm SO(10)}_L \times {\rm SO(10)}_R$ 
model \cite{Koide-so10}, where  ${\rm SO(10)}_L \times {\rm SO(10)}_R$ 
is broken into $[{\rm SU(2)}\times {\rm SU(2)}' \times {\rm SU(4)}]_L
\times [{\rm SU(2)}\times {\rm SU(2)}' \times {\rm SU(4)}]_R$.)
Then, the gauge structure above $\mu=\Lambda_S$ is different from
the present model, so that the evolutions will be also different
from the present results.
(Of course, the evolutions below $\mu=\Lambda_S$ are still the same
as those in the present paper.)
It is likely that the gauge structure above $\mu=\Lambda_S$ is
different from the present model.
Our next task is to investigate what gauge structure above
$\mu=\Lambda_S$ is promising for a unified description of
the quark and lepton masses and mixings.

\acknowledgements{
The authors  thank N.~Okamura and A.~Ghosal for their helpful comments
on the SUSY version of the universal seesaw model.
They also thank T.~Matsuki for his helpful comments on the treatment of
the renomalization group equations. 
}
\vspace{0.2in}
\appendix
{\large\bf Appendix}\\

\renewcommand{\theequation}{A.\arabic{equation}}
\setcounter{equation}{0}

In Secs. 4 and 5, the coefficients of RGE (3.6), (3.12) and (3.13) 
have been given only for the case of non-SUSY scenario with one 
SU(2)-doublet Higgs scalar.  
In the present Appendix, we give the coefficients of RGE in the 
minimal SUSY scenario.  

\vspace{2mm}
{\bf [Range III]}
\begin{eqnarray}
T_A^u&=&T_A^{\nu}=3{\rm Tr}(Y_A^uY_A^{u\dagger})
+{\rm Tr}(Y_A^{\nu}Y_A^{\nu\dagger}), \nonumber\\
T_A^d&=&T_A^e=3{\rm Tr}(Y_A^dY_A^{d\dagger})
+{\rm Tr}(Y_A^eY_A^{e\dagger}), 
\end{eqnarray}
\begin{eqnarray}
&&G_A^u={\frac{13}{6}}g_1^2+3g_2^2+{\frac{16}{3}}g_3^2+g_X^2, \nonumber\\
&&G_A^d={\frac{7}{6}}g_1^2+3g_2^2+{\frac{16}{3}}g_3^2+g_X^2, \nonumber\\
&&G_A^{\nu}={\frac{9}{6}}g_1^2+3g_2^2+g_X^2, \\
&&G_A^e={\frac{27}{6}}g_1^2+3g_2^2+g_X^2, \nonumber
\end{eqnarray}
\begin{eqnarray}
&&H_A^u=3Y_A^uY_A^{u\dagger}+Y_A^dY_A^{d\dagger},\nonumber\\
&&H_A^d=3Y_A^dY_A^{d\dagger}+Y_A^uY_A^{u\dagger},\nonumber\\
&&H_A^{\nu}=3Y_A^{\nu}Y_A^{\nu\dagger}+Y_A^eY_A^{e\dagger},\\
&&H_A^e=3Y_A^eY_A^{e\dagger}+Y_A^{\nu}Y_A^{\nu\dagger},\nonumber
\end{eqnarray}
\begin{eqnarray}
&&T_S^u=T_S^{\nu}=3{\rm Tr}(Y_S^uY_S^{u\dagger})+{\rm Tr}
(Y_S^{\nu}Y_S^{\nu\dagger}), \nonumber\\
&&T_S^d=T_S^e =3{\rm Tr}(Y_S^dY_S^{d\dagger})+
{\rm Tr}(Y_S^eY_S^{e\dagger}), 
\end{eqnarray}
\begin{eqnarray}
&&G_S^u={\frac{8}{3}}g_1^2+{\frac{16}{3}}g_3^2+3g_X^2, \nonumber\\
&&G_S^d={\frac{2}{3}}g_1^2+{\frac{16}{3}}g_3^2+3_X^2,\nonumber\\
&&G_S^{\nu}=3g_X^2, \\
&&G_S^e={\frac{18}{3}}g_1^2+3g_X^2,\nonumber
\end{eqnarray}
\begin{equation}
H_S^f=2Y_S^f Y_S^{f\dagger},
\end{equation}
where $A=L,R$ and $f=u,d,\nu,e$.  

\vspace{2mm}
{\bf [Range II]}
\begin{equation}
T_A^u= 3{\rm Tr}(Y_A^uSY_A^{u\dagger}),
\end{equation}
\begin{equation}
G_A^u={\frac{13}{6}}g_1^2+3g_{2A}^2+{\frac{16}{3}}g_3^2,
\end{equation}
\begin{equation}
H_A^u=3Y_A^uSY_A^{u\dagger},
\end{equation}
\begin{eqnarray}
&&T_K^u=3{\rm Tr}(Y_L^uSY_L^{u\dagger}+Y_R^uSY_R^{u\dagger}),\nonumber\\
&&T_K^d=T_K^e=0,
\end{eqnarray}
\begin{equation}
G_K^u=G_K^d=G_K^e={\frac{9}{2}}g_1^2+{\frac{9}{2}}
(g_{2L}^2+g_{2R}^2),
\end{equation}
\begin{eqnarray}
&&H_{KA}^u=H_{KA}^d={\frac{2}{3}}Y_ASY_A^{u\dagger},\nonumber\\
&&H_{KA}^e=0,
\end{eqnarray}
\begin{equation}
T_K^{\nu}=6{\rm Tr}(Y_L^uSY_L^{u\dagger}),
\end{equation}
\begin{equation}
G_K^{\nu}={\frac{9}{2}}g_1^2+9g_{2L}^2,
\end{equation}
\begin{equation}
H_{KL}^{\nu}=0,
\end{equation}
where $A=L,R$.  

\vspace{2mm}
{\bf [Range I]}
\begin{equation}
T_L^u=3{\rm Tr}(Y_L^uSY_L^{u\dagger}),
\end{equation}
\begin{equation}
G_L^u=\frac{13}{15}g_1^2 +3g_{2A}^2+{\frac{16}{3}}g_3^2 ,
\end{equation}
\begin{equation}
H_L^u=3Y_uSY_u^{\dagger},
\end{equation}
\begin{eqnarray}
&&T_K^u=3{\rm Tr}(Y_L^uSY_L^{u\dagger}),\nonumber\\
&&T_K^d=T_K^e=0,
\end{eqnarray}
\begin{equation}
G_K^u=G_K^d=G_K^e={\frac{9}{10}}g_1^2+{\frac{9}{2}}g_{2L}^2,
\end{equation}
\begin{eqnarray}
&&H_{KL}^u=H_{KL}^d={\frac{2}{3}}Y_L^uSY_L^{u\dagger},\nonumber\\
&&H_{KR}^u=H_{KR}^d=0,\\
&&H_{KL}^e=H_{KR}^e=0,\nonumber
\end{eqnarray}
\begin{equation}
T_K^{\nu}=6{\rm Tr}(Y_L^uSY_L^{u\dagger}),
\end{equation}
\begin{equation}
G_K^{\nu}={\frac{9}{10}}g_1^2 + 9g_{2L}^2,
\end{equation}
\begin{equation}
H_{KL}^{\nu}=0.
\end{equation}


\end{multicols}

\newpage
\vspace{5mm}
\mediumtext
\begin{table}
\caption{Quantum numbers of the fermions $f$ and $F$ 
and Higgs scalars $\phi_L$, $\phi_R$ and $\Phi$ for ${\rm SU(2)}_L \times 
{\rm SU(2)}_R \times {\rm U(1)}_{LR} \times {\rm U(1)}_{X}$.
}
\label{T-qn}

\vglue.1in
\begin{tabular}{|c|cccc|c|cccc|} \hline
  & $I_3^L$ & $I_3^R$ & $\frac{1}{2} Y_{LR}$ & $X$ &  
 & $I_3^L$ & $I_3^R$ & $\frac{1}{2} Y_{LR}$ & $X$  \\ \hline
$u_L$ & $+\frac{1}{2}$ & 0 & $+\frac{1}{6}$ & 0 &
$u_R$ & 0 & $+\frac{1}{2}$  & $+\frac{1}{6}$ & 0 \\ 
$d_L$ & $-\frac{1}{2}$ & 0 & $+\frac{1}{6}$ & 0 &
$d_R$ & 0 & $-\frac{1}{2}$  & $+\frac{1}{6}$ & 0 \\ \hline
$\nu_L$ & $+\frac{1}{2}$ & 0 & $-\frac{1}{2}$ & 0 &
$\nu_R$ & 0 & $+\frac{1}{2}$  & $-\frac{1}{2}$ & 0  \\
$e_L$ & $-\frac{1}{2}$ & 0 & $-\frac{1}{2}$ & 0 &
$e_R$ & 0 & $-\frac{1}{2}$  & $-\frac{1}{2}$ & 0 \\ \hline
$U_L$ & 0 & 0 & $+\frac{2}{3}$  & $+\frac{1}{2}$ &
$U_R$ &  0 & 0 & $+\frac{2}{3}$ & $-\frac{1}{2}$ \\ 
$D_L$ & 0 & 0 & $-\frac{1}{3}$ & $-\frac{1}{2}$ &
$D_R$ & 0 & 0 & $-\frac{1}{3}$ & $+\frac{1}{2}$ \\ \hline
$N_L$ & 0 & 0 & 0 & $+\frac{1}{2}$  &
$N_R$ &  0 & 0 & 0  & $-\frac{1}{2}$  \\
$E_L$ & 0 & 0 & $-1$ & $-\frac{1}{2}$  &
$E_R$ & 0 & 0 & $-1$ & $+\frac{1}{2}$  \\  \hline
$\phi_L^+$ & $+\frac{1}{2}$ & 0 & $+\frac{1}{2}$ & $-\frac{1}{2}$  &
$\phi_R^+$ & 0 & $+\frac{1}{2}$ & $+\frac{1}{2}$ & $+\frac{1}{2}$  \\
$\phi_L^0$ & $-\frac{1}{2}$ & 0 & $+\frac{1}{2}$ & $-\frac{1}{2}$  &
$\phi_R^0$ & 0 & $-\frac{1}{2}$ & $+\frac{1}{2}$ & $+\frac{1}{2}$  
\\ \hline
$\Phi^0$ &  0 & 0 & 0  & $ +1$ &
         &    &   &     &     \\ \hline
\end{tabular}

\end{table}

\widetext
\begin{table}
\caption{The squared mass difference $\Delta m_{ij}^2 = 
m_{\nu i}^2-m_{\nu j}^2$.
The values of the input parameters are the same as in 
Figs.~9 and 10.
The absolute values of $|\Delta m^2_{ij}|$ should not be taken
rigidly, because we can adjust those by the value of $\Lambda_S$.
}
\label{T-dm}

\vglue.1in
\begin{tabular}{|c|ccc|ccc|} \hline
  & \multicolumn{3}{c|}{non-SUSY model} & 
\multicolumn{3}{c|}{SUSY model} \\ \hline
  & at $\mu=\Lambda_L$ & at $\mu=\Lambda_S$ & at $\mu=\Lambda_X$ &
 at $\mu=\Lambda_L$ & at $\mu=\Lambda_S$ & at $\mu=\Lambda_X$ \\ \hline
$|\Delta m_{32}^2|$ [eV$^2$] & $2.39\times 10^{-3}$ & 
$9.32\times 10^{-3}$ & $3.49\times 10^{-4}$ & 
$2.72\times 10^{-3}$ & $2.51\times 10^{-3}$ & $4.08\times 10^{-4}$ 
\\ \hline  
$|\Delta m_{21}^2|$ [eV$^2$] & $1.83\times 10^{-2}$ & 
$7.15\times 10^{-1}$ & $7.67\times 10^{-2}$ &
$1.35\times 10^{-2}$ & $1.25\times 10^{-2}$ & $1.01\times 10^{-2}$
\\ \hline
$|\Delta m_{32}^2/\Delta m_{21}^2|$ & $1.30\times 10^{-1}$ & 
$1.30\times 10^{-1}$ & $4.56\times 10^{-3}$ & 
$2.02\times 10^{-1}$ & $2.02\times 10^{-1}$ & $4.04\times 10^{-2}$
\\ \hline
$|V_{23}|^2$ & $0.485$ & $0.485$ & $0.500$ &
$0.478 $ & $0.478$ & $0.500$ \\ \hline
$|V_{12}|^2$  & $0.00484$ & $0.00484$ & $0.00471$ &
$0.00492$  & $0.00492$ & $0.00466$ \\ \hline
\end{tabular}

\end{table}

\newpage

\begin{multicols}{2}
\narrowtext

\narrowtext
\narrowtext
\narrowtext
\narrowtext


\narrowtext


%



%


\end{multicols}


\vspace{.3in}

\begin{thebibliography}{99}
%
\bibitem{q-evol} S.~R.~Ju\'{a}rex W., S.F.~Herrera H., 
P.~Kielanowski and G.~Mora H., talk presented at the {\it IX
Mexican School on Particles and Fields}, August 9-19, 2000, 
Pue., Mexico. hep-ph/0009148 (2000); 
C.~R.~Das and M.~K.~Parida,  NEHU/PHYS-MP-03/2000, 
hep-ph/0010004 (2000).
%
\bibitem{nu-evol} P.~H.~Chankowski and Z.~Pluciennik,
Phys.~Lett. {\bf B316}, 312 (1993);
K.~S.~Babu, C.~N.~Leung and J.~Pantaleone,
Phys.~Lett. {\bf B319}, 191 (1993).
For a very recent work, for example, see, 
Amol S.~Dighe and A.~S.~Joshipura, CERN-TH/2000-301,
hep-ph/0010079 (2000).
%
\bibitem{USM} Z.~G.~Berezhiani, Phys.~Lett.~{\bf 129B}, 99 (1983);
Phys.~Lett.~{\bf 150B}, 177 (1985);
D.~Chang and R.~N.~Mohapatra, Phys.~Rev.~Lett.~{\bf 58},1600 (1987); 
A.~Davidson and K.~C.~Wali, Phys.~Rev.~Lett.~{\bf 59}, 393 (1987);
S.~Rajpoot, Mod.~Phys.~Lett. {\bf A2}, 307 (1987); 
Phys.~Lett.~{\bf 191B}, 122 (1987); Phys.~Rev.~{\bf D36}, 1479 (1987);
K.~B.~Babu and R.~N.~Mohapatra, Phys.~Rev.~Lett.~{\bf 62}, 1079 (1989); 
Phys.~Rev. {\bf D41}, 1286 (1990); 
S.~Ranfone, Phys.~Rev.~{\bf D42}, 3819 (1990); 
A.~Davidson, S.~Ranfone and K.~C.~Wali, 
Phys.~Rev.~{\bf D41}, 208 (1990); 
I.~Sogami and T.~Shinohara, Prog.~Theor.~Phys.~{\bf 66}, 1031 (1991);
Phys.~Rev. {\bf D47}, 2905 (1993); 
Z.~G.~Berezhiani and R.~Rattazzi, Phys.~Lett.~{\bf B279}, 124 (1992);
P.~Cho, Phys.~Rev. {\bf D48}, 5331 (1994); 
A.~Davidson, L.~Michel, M.~L,~Sage and  K.~C.~Wali, 
Phys.~Rev.~{\bf D49}, 1378 (1994); 
W.~A.~Ponce, A.~Zepeda and R.~G.~Lozano, 
Phys.~Rev.~{\bf D49}, 4954 (1994).
%
\bibitem{KFzp} Y.~Koide and H.~Fusaoka, Z.~Phys. {\bf C71}, 459 (1996); 
Prog.~Theor.~Phys. {\bf 97}, 459 (1997).
%
\bibitem{KFptp} Y.~Koide and H.~Fusaoka, 
Prog.~Theor.~Phys. {\bf 97}, 459 (1997).
%
\bibitem{Morozumi} T.~Morozumi, T.~Satou, M.~N.~Rebelo and 
M.~Tanimoto, Phys.~Lett. 
{\bf B410}, 233 (1997).
%
\bibitem{CKM} N.~Cabibbo, Phys.~Rev.~Lett.~{\bf 10}, 531 (1996); 
M.~Kobayashi and T.~Maskawa, Prog.~Theor.~Phys.~{\bf 49}, 652 (1973).
%
\bibitem{Koide-mpl} Y.~Koide, Mod.~Phys.~Lett. {\bf 22}, 2071 (1993).
%
\bibitem{evol-USM} Y.~Koide, Phys.~Rev. {\bf D58}, 053008 (1998).
%
\bibitem{qmass} H.~Fusaoka and Y.~Koide, Phys.~Rev. {\bf D57}, 3986
(1998).

%
\bibitem{PDG00} Particle data group, Eur.~Phys.~Jour. {\bf C15}, 1 (2000).
%
\bibitem{MNS} Z.~Maki, M.~Nakagawa and S.~Sakata, Prog.~Theor.~Phys.
{\bf 28}, 870 (1962).
%
\bibitem{Zee} A.~Zee, Phys.~Lett. {\bf 93B}, 389 (1980).
%
\bibitem{mixing-Zee} For example, see,
A.~Yu.~Smirnov and M.~Tanimoto, Phys.~Rev. {\bf D55}, 1665 (1997);
C.~Jarlskog, M.~Matsuda, S.~Skadhauge, M.~Tanimoto, 
Phys.~Lett. {\bf B449}, 240 (1999).
%
\bibitem{nu-Koide}
Y.~Koide, Mod. Phys.~Lett. {\bf 11}, 2849 (1996);
Phys.~Rev. {\bf D57}, 5836 (1998).
%
\bibitem{nu-atm} Y.~Fukuda {\it et al.}, Phys.~Lett. {\bf B335}, 
237 (1994);
Super-Kamiokande collaboration, Y.~Fukuda, {\it et. al.},
Phys.~Rev.~Lett. {\bf 81}, 1562 (1998);
H.~Sobel, Talk presented at {\it Neutrino 2000},
Sudbury, Canada, June 2000 (http://nu2000.sno.laurentian.ca/).
%
\bibitem{dem-nu-evol} N.~Haba, Y.~Matsui, N.~Okamura and T.~Suzuki,
Phys.~Lett. {\bf B489}, 184 (2000).
%
\bibitem{nu-solar} Y.~Suzuki, Talk presented at {\it Neutrino 2000},
Sudbury, Canada, June 2000 (http://nu2000.sno.laurentian.ca/).
Also see, M.~Gonzalez-Garcia, Talk presented at {\it Neutrino 2000},
Sudbury, Canada, June 2000 (http://nu2000.sno.laurentian.ca/).
\bibitem{nu-DUSM} Y.~Koide and H.~Fusaoka, Phys.~Rev. {\bf D59},
053004 (1999); Y.~Koide and A.~Ghosal, Phys.~Lett. {\bf B488},
344 (2000).
%
%
\bibitem{Koide-so10} Y.~Koide, Phys.~Rev. {\bf D61},
035008 (2000).
\end{thebibliography}
\end{document}